\documentclass[12pt]{article}
\usepackage{graphicx}
\usepackage{setspace}
\usepackage{rotating}
\usepackage{epsf, color, verbatim}
\usepackage{amssymb,amsmath}
\usepackage{soul}
\usepackage{color}
\usepackage[usenames,dvipsnames]{xcolor}
\usepackage{caption} 
\usepackage{booktabs}
\usepackage{tabulary}
\usepackage{biblatex} 
\newcolumntype{K}[1]{>{\centering\arraybackslash}p{#1}}

\newcommand{\bfmath}[1]{\mbox{\boldmath$#1$\unboldmath}}

\title{\textbf{Statistical Monitoring of the Covariance Matrix in Multivariate Processes: A Literature Review}}

\usepackage{authblk}
\title{\textbf{Statistical Monitoring of the Covariance Matrix in Multivariate Processes: A Literature Review}}
\author[1]{\small{Mohsen Ebadi}}
 \author[1]{\small{Shoja'eddin Chenouri}}
  \author[2]{\small{Dennis K. J. Lin}} 
  \author[1]{\small{Stefan H. Steiner}}
  \affil[1]{ {\small \it  Department of Statistics and Actuarial Science, University of Waterloo, Canada}} 
   \affil[2]{ {\small \it  Department of Statistics, Purdue University, West Lafayette, IN, USA}}
\date{}

\begin{document}
\maketitle
\pagestyle{myheadings}
\begin{abstract}
Monitoring several correlated quality characteristics of a process is common in modern manufacturing and service industries. Although a lot of attention has been paid to monitoring the multivariate process mean, not many control charts are available for monitoring the covariance matrix. This paper presents a comprehensive overview of the literature on control charts for monitoring the covariance matrix in a multivariate statistical process monitoring (MSPM) framework. It classifies the research that has previously appeared in the literature. We highlight the challenging areas for research and provide some directions for future research.
\end{abstract}
{\fontsize{10}{0}\selectfont KEYWORDS: Covariance matrix; Multivariate Control Charts; Phase I and Phase II; Statistical Process Monitoring (SPM).\\
------------------------------------------------------------------------------------------------ \\
%
\fontfamily{time}\selectfont }
\section{Introduction}
Statistical process monitoring (SPM) is an effective and widely used method for quality improvement of manufacturing and service products. Among the tools of SPM, control charts are the most powerful techniques. In many situations, the goal is to monitor many inter-related characteristics of a process simultaneously. For example, Huwang et al. (2007) provided an application of wafer manufacturing in the chemical mechanical planarization process from the semiconductor industry and Maboudou-Tchao and Agboto (2013) explained an example from healthcare where they monitored the vital signs of a patient, such as systolic blood pressure, heart rate, and body temperature in an intensive care unit (ICU). These types of examples have motivated the development of many multivariate statistical process monitoring (MSPM) control charts for different features of the underlying multivariate processes, such as the mean vector, covariance matrix, shape matrix, etc. 

Since the publication of the pioneering paper of Hotelling (1947) on multivariate control charts, numerous techniques have been introduced. Jackson (1985), Lowry and Montgomery (1995), and Bersimis et al. (2007) have provided excellent reviews of the literature, but their focus is mainly on detecting mean shifts. Monitoring the covariance structure of multivariate processes is also of a great importance, but has received much less attention in the literature. In addition, common methods of multivariate quality monitoring like that of Hotelling, which is based on quadratic forms, confound mean shifts with scale shifts. This implies that upon receiving a signal from a control chart, an extensive analysis is needed to determine the nature of the shift.  

Monitoring covariance or dispersion matrices is a much more challenging problem than detecting mean shifts for several reasons. First, it is not clear how best to summarize the entries of a covariance matrix. Second, for $p\geq3$ quality characteristics, the number of unknown parameters to be estimated is much larger than with a mean vector. These types of considerations have manifested in methodologies that are inherently more diverse than those for monitoring the mean and therefore are not easily accessible for practitioners. Hence, it is useful to summarize and discuss the diverse literature on the topic in a structured and easily accessible manner. Yeh et al. (2006) reviewed control chart methodologies for covariance matrix monitoring developed between 1990 and 2005. However, since then there have been many advances on several fronts. The purpose of this paper is to update the existing literature of this area by focusing on developments published since 2006. To do this, in a systematic way, we classify the research into distinctive categories. The key idea in the current review paper is to highlight critical issues and provide guidelines for selecting suitable control charts in specific settings.  

This review mainly focuses on Phase II control charts designed for multivariate normal processes with independent subgroups of observations, but also considers related topics, such as diagnostic procedures, robust control charts, and Phase I analysis methods. The remainder of this paper is organized as follows. Section 2 deals with definitions, notation, and problem settings and describes the classification approach we adopt for grouping the relevant publications. Section 3 discusses the methods available in the literature for monitoring the covariance matrix when the sample size is greater than or equal to the number of process variables, while Section 4 presents the relevant work on monitoring the covariance matrix when sample size is less than the number of  variables. Section 5 reviews the control charts for simultaneously monitoring the mean vector and covariance matrix. In Section 6, other developments related to covariance monitoring such as interpretation of an out-of-control (OC) signal, robust control charts, and Phase I analysis are given. Finally, concluding remarks and recommendations for future research are presented in Section 7. 

\section{Problem definition and a literature classification scheme}
Statistical process monitoring contains two main phases: Phase I and Phase II (Woodall and Montgomery 1999). In Phase I, historical observations are analyzed in order to determine whether or not a process is in-control and then, to estimate the process parameters of the in-control process. The emphasis in Phase I is on finding good parameter estimates of a model for the in-control process. There are two dominant approaches in the literature to achieve this. The most popular approach is to first clean the Phase I data to remove outliers and then calculate the classical efficient estimates based on the cleared data. The second approach is to calculate robust estimates of the in-control process parameters. Here, robustness refer to "distribution-robustness" and robustness against contaminated data or outliers generated due to the violation of the distributional assumptions. The Phase II data analysis, however, consists of using the estimates obtained in Phase I and the derived control limits to examine whether or not the process remains in-control. The emphasis here is quick detection of an out-of-control situation. Typically, Phase I analysis is an iterative task and a lot of process understanding and possibly improvements are needed in the transition from Phase I to Phase II (Woodall 2000).

Consider a multivariate process with $p$ quality characteristics $\mathbf{X}=(X_1,\,X_2,\,\dots,\,X_p)'$ to be monitored simultaneously.  Under an in-control situation, we assume that the process follows a multivariate normal distribution with mean vector $\bfmath{\mu}_{_0}=({\mu}_{_{0\,1}},\,\dots ,\,{\mu}_{_{0\,p}})'$ and covariance matrix $\bfmath{\Sigma}_{_0}$. Let $\bfmath{\sigma}_{_0}=({\sigma}_{_{0\,1}},\,\dots,\,{\sigma}_{_{0\,p}})'$ denote the vector of in-control standard deviations of these $p$ variables obtained from the diagonal elements of $\bfmath{\Sigma}_{_0}$. We assume that independent sample sets $\mathbb{X}_{1},\,\mathbb{X}_{2},\,\dots $, each with cardinality $n$ in the form of $\mathbb{X}_i=\lbrace \mathbf{X}_{i\,1},\,\dots,\,\mathbf{X}_{i\,n} \rbrace $, are taken periodically from the process. In sampling epoch $i$, we denote the $j$th observation of the $k$th quality characteristic  $X_{i\,j\,k},$ where $k=1,\,\dots,\,p$, $j=1,\,\dots,\,n,$ and $i=1,\,2,\,\dots\,$. Suppose in Phase I, we have collected $N$ such sample sets and we estimated the parameters $\bfmath{\mu}_{_0}$ and $\bfmath{\Sigma}_{_0}$ by the sample mean vector and sample covariance matrix of the combined $N\,n$  observations, respectively, which are defined as: 
\begin{equation}
\overline{\overline{\mathbf{X}}}=\frac{\sum_{i=1}^{N}{\sum_{j=1}^{n}{\mathbf{X}_{i\,j}}}}{Nn}, \qquad 
\overline{\mathbf{S}}=\frac{\sum_{i=1}^{N}{\sum_{j=1}^{n}{(\mathbf{X}_{i\,j}}-\overline{\mathbf{X}}_i){(\mathbf{X}_{i\,j}}-\overline{\mathbf{X}}_i)^{\prime}}}{N\,(n-1)},
\end{equation}
where $\overline{\mathbf{X}}_i={\sum_{j=1}^{n}{\mathbf{X}_{i\,j}}}/n$. The existing literature on control charts for monitoring covariance matrices is mainly motivated by testing two-sided or one-sided hypotheses testing problems such as:
\begin{equation}\label{2WayTest}
\bfmath{\Sigma}=\bfmath{\Sigma}_0 \qquad \text{ versus }\qquad \bfmath{\Sigma}\neq\bfmath{\Sigma}_0\,,
\end{equation}
\begin{equation}\label{twosided}
\bfmath{\Sigma}=\bfmath{\Sigma}_0 \qquad \text{ versus }\qquad \bfmath{\Sigma}\succcurlyeq \bfmath{\Sigma}_0\,, \qquad \text{ or } \qquad \bfmath{\Sigma}=\bfmath{\Sigma}_0 \qquad \text{ versus }\qquad \bfmath{\Sigma}\preccurlyeq \bfmath{\Sigma}_0\,,
\end{equation}
where the notations $\bfmath{\Sigma}\succcurlyeq \bfmath{\Sigma}_0$ and $\bfmath{\Sigma}\preccurlyeq \bfmath{\Sigma}_0$ mean that $\bfmath{\Sigma}-\bfmath{\Sigma}_0$ and $\bfmath{\Sigma}_0-\bfmath{\Sigma}$ are positive semidefinite (PSD) matrices, respectively. On the other hand, the methodology for simultaneous monitoring of a mean vector and a covariance matrix is often inspired by the test of hypothesis: 
\begin{equation}
\bfmath{\mu}=\bfmath{\mu}_0\,\quad\textrm{and} \quad \bfmath{\Sigma}=\bfmath{\Sigma}_0 \qquad \text{ versus }\qquad \bfmath{\mu}\neq \bfmath{\mu}_0\quad\textrm{or} \quad \bfmath{\Sigma}\neq\bfmath{\Sigma}_0\,.
\end{equation}

By surveying some existing multivariate methodologies for two-sided hypothesis test of covariance matrices, Alt (1985) introduced a number of control charts for monitoring covariance matrices of multivariate processes. Note that it is usually much more important to detect a dispersion increase than a decrease, which means that methods based on a one-sided test are more appropriate in this context than two-sided tests. However, as stated by Yen and Shiau (2010), for two multivariate processes
perhaps there is no way to define precisely which has larger (or smaller) covariance matrix, but use of one-sided tests is just a “natural” way for this objective.

As mentioned above, Phase II monitoring of a process covariance matrix has been the main concern of the existing articles in this research area. Some of the existing methodologies are direct extensions of  univariate control charts to the multivariate framework, while others are based on different approaches. In this review paper, we divide the existing literature into four categories.  Below, we provide some explanations on how these categories are formed:
\begin{itemize}
\item[a) ] {\bf Monitoring covariance matrix alone, when $n\geq{p}$ :}\\
This case has been explored the most among all four categories. The literature assumes that the rational group size $n$ is large enough so that the sample covariance matrices have full rank. 
\item[b) ] {\bf Monitoring covariance matrix alone, when $n< p$ :}\\
When the number of variables $p$ is large, collecting samples with more observations than variables is sometimes not practical. As an extreme case, there are some situations where due to the expense, only individual observations can be sampled for subgroups and therefore the usual methodologies based on the sample covariance matrix are not applicable. 

\item[(c) ] {\bf Simultaneous monitoring of mean and covariance:}\\
In multivariate setting, process changes may happen in many ways. For example, one variable or characteristic may have shift in the mean while another variable may have a shift in its variability. It is also possible that correlations between variables may change in some more complicated way. However, this possibility has received little attention in the literature. Recent papers have mainly focused on procedures for simultaneous monitoring of the mean vector and the covariance matrix. 
\item[(d) ] {\bf Other related topics to covariance monitoring:}\\
This category contains the other topics related to the monitoring of a covariance matrix in multivariate processes such as diagnostic procedures, robust approaches and Phase I analysis.
\end{itemize}
A commonly used performance measure in the literature to compare the performance of different control charts is Average Run Length (ARL), which is the average number of points plotted on a control chart until an out-of-control signal is observed. In this paper, ARLs under in-control and out-of-control situations are denoted by $\rm{ARL}_0$ and $\rm{ARL}_1$, respectively. Typically, a long $\rm{ARL}_0$ is preferred, while a short $\rm{ARL}_1$ is desired. Variants of ARL, like Average Time to Signal (ATS), are also used in the literature. In what follows, we shall note that some methods reviewed in this paper can potentially belong to more than one category. For example, the methods used for the case $n< p$ can sometimes be extended to the case $n\geq p$. In the next four sections, the main idea of related papers to each category will be discussed.
\section{Monitoring covariance matrix alone, when $n\geq{p}$}
As previously mentioned, the literature for the case of $n\geq{p}$ is abundant. Generally, we summarize  the information in multivariate data by means of a univariate statistic. Shewhart-type control charts for monitoring covariance matrices are often constructed based on either the determinant of the sample covariance matrix, known as the sample generalized variance and denoted by $|\,\mathbf{S}\,|$, or by the trace of the sample covariance matrix, which is the sum of the variances of the variables and denoted by ${\rm tr}(\mathbf{S})$. However, both $|\,\mathbf{S}\,|$ and ${\rm tr}(\mathbf{S})$ fail to capture certain important features. Since the determinant $|\,\mathbf{S}\,|$ is the product of the eigenvalues, it is incapable of detecting covariance matrix changes when some aspects of the variability increase while others decrease. On the other hand, $\rm tr(\mathbf{S})$, being the sum of marginal variances, does not take into account the covariances. 

Alt (1985) used both of these aforementioned quantities and proposed a few methods for Phase II monitoring of covariance matrices. Alt (1985) and Alt and Smith (1988) adapted a multivariate control chart based on the likelihood ratio test (LRT) statistic for the one-sample test of the covariance matrix in \eqref{2WayTest} and defined a charting statistic of the form
\begin{equation}
R_{i}=-(n-1)\left[p+\ln\frac{|\mathbf{S}_{i}|}{|\bfmath{\Sigma}_{0}|}-{\rm tr}(\bfmath{\Sigma_{0}^{-1}}\mathbf{S}_{i})\right],  \quad  
  i = 1,\,2,\, \dots\,, 
\end{equation}
where $\mathbf{S}_{i}$ is the $p\times p$ sample variance-covariance matrix of the rational subgroup $i$ with size $n$. 
Asymptotic upper control limits (UCL) for this Shewhart-type chart are given by the upper percentiles of $\chi^{2}$ distribution with degrees of freedom $p\,(p+1)/2$ for large $n$. This method can detect departures in any direction from the in-control (IC) setting. Another method proposed by Alt (1985) is to monitor the square root of $|\mathbf{S}|$, where the control limits were only given for the case of $p=2$ obtained from the probability 
\begin{equation}
\Pr\left[\frac{\chi^{2}_{2n-4,(1-\alpha/2)}}{2(n-1)}\leq \sqrt{\frac{\mid \mathbf{S}\mid}{\mid \bfmath{\Sigma}_{0}\mid}}\leq\frac{\chi^{2}_{2n-4,\alpha/2}}{2(n-1)}\right ]=1-\alpha\,.
\end{equation}
The exact distribution of $|\mathbf{S}|$ for $p>2$ is not known. Hence, for higher dimensions, Alt (1985) proposed an alternative method to monitor $|\mathbf{S}|$ by assuming that the probability distribution of $|\mathbf{S}|$ is mostly contained in the interval $\rm E(|\mathbf{S}|)\pm 3\sqrt{\rm Var(|\mathbf{S}|)}$. 
Since then, several publications have discussed extensions and properties of $|\mathbf{S}|$ chart. Two recent publications are Dogu and Kocakoc (2011) and Lee and Khoo (2018).
Ghute and Shirke (2008) proposed a synthetic charting approach for monitoring a covariance matrix by combining the $|\mathbf{S}|$ chart of Alt (1985) and what they call the Conforming Run Length (CRL) chart. This approach is a two-step procedure. For a given rational subgroup, if $|\mathbf{S}|$ chart triggers an out-of-control alarm, that rational subgroup is declared as nonconforming, otherwise it is declared as conforming. The CRL is the number of conforming rational subgroups between two consecutive nonconforming rational subgroups including the end nonconforming subgroup. Let $Y_{r}$ be the $r$th ($r=1,2,...$) CRL. The procedure declares the process out-of-control, if $Y_{r}$ is less than or equal to a predefined value. Gadre (2014) proposed group runs $|\mathbf{S}|$ control charts as an extension of the methodology in Ghute and Shirke (2008). The group runs chart declares a process out-of-control, if $Y_{1}\leq {L}$ or, if for some $r>1,\,Y_{r}$ and $Y_{(r+1)}\leq{L}$ for the first time for some predefined value $L$. More work on improving the performance of synthetic $|\mathbf{S}|$ control charts for covariance matrix can be found in Lee and Khoo (2015, 2017a, and 2017b).

Costa and Machado (2009) used the statistic VMAX, the maximum of the marginal sample variances of standardized variables from rational subgroups where standardization is done using the marginal in-control mean and standard deviations. To be more specific, for the general multivariate case at sampling epoch $i$, the VMAX statistic is defined as
\begin{equation}
\textrm{VMAX}_{i}=\underset{k=1,\,\dots,\, p}{\textrm{max}}\Bigg\{\frac{\sum_{j=1}^{n}{(X_{i\,j\,k}-{\mu}_{_{0\,k}})^{2}}}{n{\sigma}_{_{0\,k}}^2}\Bigg\}.
\end{equation}
Although the marginal variances can be also computed when $n<p$, the larger sample size can provide better estimation and Costa and Machado (2009) only included the case $n>{p}$ in their numerical studies. Once a signal occurred, the responsible variable for the out-of-control situation is identified as the one with a sample variance larger than the upper control limit.  It is argued that $\textrm{VMAX}$ has the advantage of faster detection of covariance shifts and is better at identifying the individual out-of-control variables in comparison to the $|\mathbf{S}|$ chart introduced in Alt (1985). However, this method is sensitive to linear dependency among the variables, as the $\rm ARL_1$ of the method increases when the correlations among the variables increase. Costa and Machado (2008), Machado et al. (2009b), Quinino et al. (2012), Costa and Neto (2017), Gadre and Kakade (2018), and Machado et al. (2018) proposed some variants of VMAX and improved it in bivariate and trivariate cases. For example, Quinino et al. (2012) introduced VMIX chart for a bivariate normal process at sampling epoch $i$ as:
\begin{equation}
\textrm{VMIX}_{i}=\frac{\sum_{j=1}^{n}X_{i\,j\,1}^{*^{2}}+\sum_{j=1}^{n}X_{i\,j\,2}^{*^{2}}}{2n},
\end{equation}
where
\begin{equation}
X_{i\,j\,1}^{*}=\frac{X_{i\,j\,1}-{\mu}_{_{0\,1}}}{{\sigma}_{_{0\,1}}}, \qquad   X_{i\,j\,2}^{*}=\frac{X_{i\,j\,2}-{\mu}_{_{0\,2}}-\rho_{_0}\,{\sigma}_{_{0\,2}}{X_{i\,j\,1}^{*}}}{{\sigma}_{_{0\,2}}\sqrt{1-\rho_{_0}^{2}}},
\end{equation}
and where $\rho_{_0}$ is the in-control correlation coefficient of the two variables. It is shown that VMIX has lower $\rm{ARL}_1$ values than VMAX for small shifts in one variance or shifts in both variances. In addition to the aforementioned bivariate charts, Cheng and Cheng (2011) proposed an artificial neural network based approach, which showed better performance than the traditional $|\mathbf{S}|$ chart proposed by Alt (1985) in detecting shifts in the covariance matrix of a bivariate process.

Although most of the techniques developed in the literature are centered on detecting either increases or decreases in dispersion, in most practical situations it is more critical to detect a process dispersion increase. Costa and Machado (2011a) introduced the RMAX statistic based on the marginal sample ranges of the quality characteristics. Let $R_{i\,k}=\max\lbrace X_{i\,1\,k},\,X_{i\,2\,k},\,\dots,\, X_{i\,n\,k}\rbrace-\min\lbrace X_{i\,1\,k},\,X_{i\,2\,k},\,\dots,\, X_{i\,n\,k}\rbrace$ denote the sample range of $k$th quality characteristic for the $i$th rational subgroup. They propose using ${\rm RMAX}_i=\max\left\lbrace R_{i\,1},\,R_{i\,2}, \,\dots,\, R_{i\,p}\right\rbrace$  for sample $i=1,\,2,\,\dots$ as a monitoring statistic. This RMAX chart signals increases in the variances and covariances. Yen and Shiau (2010) proposed a control chart for detecting a dispersion increase, i.e.  $\bfmath{\Sigma}-\bfmath{\Sigma}_{_0}$ is positive semidefinite, based on the LRT statistic of a one-sided test. For the rational subgroup of $i$, define 
\begin{equation}
\psi_{_i}=
\left\{
	\begin{array}{ll}
		\prod\limits_{k=1}^{p^*}\left\lbrace d_{ik}\exp\left[-(d_{ik}-1)\right]\right\rbrace^{\frac{n}{2}}  & \mbox{for }  p^{*}>0 \\
		1 & \mbox{for } p^{*}=0\, ,
	\end{array}
\right.
\end{equation}
where $d_{i1}\geq d_{i2}\geq ...\geq d_{ip}> 0$ are the roots of the equation $\mid \mathbf{S}_{i}-d\,\bfmath{\Sigma}_{_0}\mid=\mathbf{0}$, and $p^*$  is the number of $d_{ik}> 1$.  The LRT statistic is $D_i=-2\,\log\psi_{_i}$ and whenever $D_i>D_\alpha$, the $(1-\alpha)$th quantile of the distribution of $D$, the process is considered to be out-of-control. 
It is important to notice that the detection of dispersion decreases can also be important and drive process improvement. See Yen et al. (2012) in which the authors used the corresponding one-sided LRT statistic for detecting a decrease in multivariate dispersion. 

The aforementioned control charts are designed for detecting shifts in any number of elements in  the covariance $\bfmath{\Sigma}$. However, in some real-world applications, it is reasonable to assume that changes are expected to only occur in a small number of entries in the covariance matrix (known as the sparsity feature) especially when $p$ is large, and thus requiring more efficient estimation methods.
Recall the in-control process $\mathbf{X}\sim N_p(\bfmath{\mu}_{_0},\,\mathbf{\Sigma}_{_0})$ and note that for the lower triangular (inverse-Cholesky root) matrix $\mathbf{A}$ with property $\mathbf{A}\,\mathbf{\Sigma}_{_0}\mathbf{A}'=\mathbf{I}_p$, we have $\mathbf{A}\,(\mathbf{X}-\bfmath{\mu}_{_0})\sim N_p(\mathbf{0},\,\mathbf{I}_p)$, where $\mathbf{I}_{p}$ is the $p\times p$ identity matrix. Thus, without loss of generality, one can assume that the in-control distribution of $\mathbf{X}$ is $N_p(\mathbf{0}, \mathbf{I}_{p})$, and if the process goes out of control, the distribution changes to $N_p(\mathbf{0}, \mathbf{\Sigma}_{_1})$ with only a few entries of $\mathbf{\Sigma}_{_1}$ different from the entries of the identity matrix $\mathbf{I}_{p}$. To incorporate this information into the design of control charts for monitoring the process variability, Li et al. (2013) proposed a penalized likelihood estimation approach. They used an $\ell_{1}$  penalty function on the precision matrix $\mathbf{\Omega}=\bfmath{\Sigma}^{-1}$ in order to force the unchanged off-diagonal elements of the precision matrix to zero. So, for sample epoch $i$, the respective estimate of the precision matrix $\widehat{\mathbf{\Omega}}_{i}$ can be defined as the solution to the following penalized likelihood function
\begin{equation}\label{PenPer1}
\widehat{\mathbf{\Omega}}_{i}(\lambda)=\underset{\mathbf{\Omega}\succ\mathbf{0}}{\operatorname{argmin}}\left\lbrace {\rm tr}(\mathbf{\Omega}\,\mathbf{S}_{i})-\ln|\mathbf{\Omega}|+\lambda\,\|\mathbf{\Omega}\|_{1}\right\rbrace,
\end{equation}
where $\|.\|_{1}$ is the matrix $\ell_{1}$-norm and $\lambda$ is the penalization or tuning parameter that can be selected to achieve different levels of sparsity for the estimated $\widehat{\mathbf{\Omega}}_{i}(\lambda)$. Having obtained $\widehat{\mathbf{\Omega}}_{i}(\lambda)$, the Penalized Likelihood Ratio (PLR) charting statistic is defined as:
\begin{equation}
\Lambda_{i}(\lambda)=\textrm{tr}(\mathbf{S}_{i})-{\rm tr}\left(\widehat{\mathbf{\Omega}}_{i}(\lambda)\mathbf{S}_{i}\right)+\ln|\widehat{\mathbf{\Omega}}_{i}(\lambda)|.
\end{equation}
The PLR chart signals an out-of-control alarm when $\Lambda_{i}(\lambda)>UCL_{\lambda}$ where $UCL_{\lambda}$ depends on the selected $\lambda$ and the desired $\rm ARL_0$. $UCL_{\lambda}$ can be obtained via a Monte Carlo simulation. 
 
In addition to the aforementioned Shewhart-type control charts developed for covariance monitoring in the case $n\geq{p}$, some memory-type control charts like the cumulative sum (CUSUM) and exponentially weighted moving average (EWMA) charts have been also considered by some authors. Memory-type control charts accumulate data from past samples and are better than Shewhart-type charts in detecting small process shifts. Healy (1987) considered the assumption that a shift in the covariance matrix happens in form of $\bfmath{\Sigma}_1=c\bfmath{\Sigma}_0$, where $c>0$ is a constant. In this formulation, when a change occurs in the process, it affects all of the variables. Then, a procedure based on a multivariate CUSUM (MCUSUM) can be developed  for detecting a shift in the covariance matrix based on the following statistic:

\begin{equation}
\xi_{i}=\textrm{max}\left(0,\xi_{i-1}+{\sum_{j=1}^{n}({\mathbf{X}_{i\,j}}-\bfmath{\mu}_{_0})'}\bfmath{\Sigma_{0}^{-1}}({\mathbf{X}_{i\,j}}-\bfmath{\mu}_{_0})-g\right), 
\end{equation}
where parameter $g=p\,n\,\textrm{log}(c)[c/(c-1)]$ and the initial $\xi_{_0}$ can take a certain value in order to make CUSUM having improved properties. The chart signals whenever $\xi_{i}>UCL$ and is also applicable for the case $n<p$. Surtihadi et al. (2004) investigated more general patterns of shifts in the covariance matrix. Three specific cases include a known scalar change only in one of the variables, an unknown scalar change only in one of the variables, and an unknown scalar change in the covariance matrix as a whole. Surtihadi et al. (2004) developed CUSUM charts as well as Shewhart-type control charts for these three cases and compared their performances through ARL comparisons. As another memory-type control chart, Machado and Costa (2008) developed EWMA charts based on the VMAX statistic to monitor the covariance matrix of bivariate processes. They also investigated the effect of double sampling (DS) on the sensitivity of a VMAX chart to moderate and small shifts. Moreover, Bartzis and Bersimis (2020) recently compared several bivariate control charts for monitoring a covariance matrix and concluded that the VMIX chart and the EWMA chart based on the VMAX statistic have the lowest $ARL_1$ (for the same $ARL_0$) between some competing methods under different simulation scenarios. Osei-Aning and Abbasi (2020) developed a set of bivariate EWMA-type charts that can be used for both normal and non-normal bivariate processes and compared their performance under different underlying distributions. Riaz et al. (2019) proposed a new multivariate dispersion chart, called a Multivariate Mixed EWMA-CUSUM (MMECD) control chart, by integrating the effects of a MEWMA and a MCUSUM chart. MMECD enhances the efficiency of memory-type control charts for monitoring a covariance matrix and a simulation indicated that it has better performance than individual MEWMA or MCUSUM charts for detecting small shifts in the process
dispersion.  
\section{Monitoring covariance matrix alone, when $n<p$}
Gathering observations with desirable rational subgroups is not always possible and there are certain processes in which only individual observations (i.e. $n=1$) can be collected. As mentioned in Section 2, the main problem in this case is that the estimated sample covariance matrix is singular and the common multivariate methods can not be used. While most existing approaches use the EWMA structure, a few Shewhart-type control charts have been introduced for the case of $n<p$. We start with presenting the Shewhart-type control charts for this case and then we mainly focus on EWMA-type control charts. Djauhari et al. (2008) introduced a new measure of multivariate variability based on vector variance (VV), which is defined as the sum of squares of all eigenvalues of covariance matrix and can be used when the covariance matrix is singular. By employing the asymptotic distribution of VV, they developed the Shewhart-type VV-chart. The resulting control chart has a better performance in terms of $ARL_1$ than control charts based on the generalized variance. Later, Djauhari et al. (2016) highlighted a drawback of the VV and generalized variance charts that the probability of false alarm (PFA) is not taken into account in the determination of their control limits. Therefore, Djauhari et al. (2016) proposed a methodology to remedy this problem, when the subgroup size is small.  Mason et al. (2009) introduced a charting approach that is in the form of the well-known Wilks' statistic. When $n<p$, the charting statistic can be expressed as a function of the ratio of the determinant of two separate estimates of the underlying covariance matrix. The numerator is the determinant of the sample covariance matrix based on the Phase I historical data and the denominator is obtained from the historical data plus the most recently observed subgroup of size $n$ in Phase II. When the test shows that a new sample does not increase the volume of the space occupied by the historical dataset, then it means that the new sample is very similar to the original sample. Otherwise, the control chart signals an out-of-control situation. Maboudou-Tchao and Agboto (2013) proposed a Shewhart-type control chart similar to that of Li et al. (2013) but designed for $n<p$. Recall the matrix $\mathbf{A}$ with property $\mathbf{A}\,\mathbf{\Sigma}_{_0}\mathbf{A}'=\mathbf{I}_p$. At the sampling epoch $i$ and subgroup $j$, one can perform the transformation $\mathbf{U}_{i\,j}=\mathbf{A}(\mathbf{X}_{i\,j}-\bfmath{\mu}_{0})$. Let $\mathbf{U}_i$ be a $n\times p$ matrix whose $j$th row is given by the row vector $\mathbf{U}_{i\,j}'$. The method of Maboudou-Tchao and Agboto (2013) uses the penalization approach given in \eqref{PenPer1} while $\mathbf{S}_{i}$ is replaced by $\mathbf{U}_{i}'\mathbf{U}_{i}$, and produces $\widehat{\mathbf{\Omega}}_{i}$ as an estimate of $\mathbf{\Sigma}^{-1}$. The charting statistic is given by 
\begin{equation}
T_{i}=\textrm{tr}(\widehat{\mathbf{\Omega}}_{i}^{-1})-\textrm{log}|\widehat{\mathbf{\Omega}}_{i}^{-1}|-p.
\end{equation}

As previously mentioned, there has been an emphasis in the literature on developing EWMA-type control charts for monitoring the covariance matrix of a process with individual observations due to their efficiency in detecting small shifts. When $n<{p}$  the determinant of the covariance-matrix estimator as a generalized variance measure is always zero. However, in these situations, the trace of the matrix could be employed as an alternative measure of generalized variance. Despite the fact that it seems more natural to use trace to monitor the process variability when the matrix is diagonal. Huwang et al. (2007) used this idea and proposed a multivariate exponentially weighted mean square (MEWMS) control chart for individual observations. They used the inverse square root of the covariance matrix and the transformation $\mathbf{Y}=\bfmath{\Sigma}_{0}^{-\frac{1}{2}}(\mathbf{X}-\bfmath{\mu}_{0})\sim N_p(\bfmath{\mu}_Y,\bfmath{\Sigma}_Y)$, whose in-control parameters are $\bfmath{\mu}_{0Y}=\mathbf{0}$ and $\bfmath{\Sigma}_{0Y}=\mathbf{I}_{p}$. Note that the root $\mathbf{A}$ in $\mathbf{A}\,\mathbf{\Sigma}_{_0}\mathbf{A}'=\mathbf{I}_p$ is not unique, so that different matrices for $\mathbf{A}$ can be used. At the sampling epoch $i$, one can define a multivariate exponentially weighted moving average statistic for individual observations as:
\begin{equation}\label{Zi}
\mathbf{Z}_{i}=\omega \mathbf{Y}_{i}\mathbf{Y}_{i}^{\prime}+(1-\omega)\mathbf{Z}_{i-1},
\end{equation}
where $0<\omega<1$ is the smoothing parameter and $\mathbf{Z}_{0}=\mathbf{Y}_1\mathbf{Y}_1^{\prime}$. When $i\geq{p}$, $\mathbf{Z}_{i}$ is a positive definite matrix with probability 1. It can be shown that when the process mean does not change, E($\mathbf{Z}_{i})=\bfmath{\Sigma}_Y$ for any given $i$, so that $\mathbf{Z}_{i}$ can be used as an estimate of $\bfmath{\Sigma}_Y$. An MEWMS control chart for covariance monitoring can be defined with time-varying control limits of the form 
\begin{equation}\label{trZi}
\textrm{E[tr}(\mathbf{Z}_{i})]\pm L\sqrt{\textrm{Var[tr}(\mathbf{Z}_{i})]}\,=p\pm L\sqrt{2p\left[\frac{\omega}{2-\omega}+\frac{2-2\omega}{2-\omega}{(1-\omega)^{2(i-1)}}\right]}.
\end{equation}
Here the value of $L$ depends on $p$, $\omega$, and $\rm ARL_0$. Notice that this procedure is sensitive to changes in the mean vector, meaning that if the mean vector of the process changes in the observation period whilst the covariance matrix remains unchanged, the procedure tends to spuriously signal a covariance change.  
To overcome this drawback, Huwang et al. (2007) proposed another procedure, called the multivariate exponentially weighted moving variance (MEWMV) chart which is insensitive to shifts in the process mean. Their proposal is to replace $\mathbf{Y}_{i}$ in (15) with its deviation from $\mathbf{P}_{i}$, in the form of 
\begin{equation}
\mathbf{V}_{i}=\omega (\mathbf{Y}_{i}-\mathbf{P}_{i})(\mathbf{Y}_{i}-\mathbf{P}_{i})^{\prime}+(1-\omega)\mathbf{V}_{i-1}\,,
\end{equation}
where $\mathbf{P}_{i}$ is an estimate of the possible mean shift at time $i$ and it is obtained from the recursion $\mathbf{P}_{i}=\nu\mathbf{Y}_{i}+(1-\nu)\mathbf{P}_{i-1}$, where $0<\nu<1$ and $\mathbf{P}_{0}=\mathbf{0}$, the matrix  $\mathbf{V}_{0}$ is defined by $\mathbf{V}_{0}=(\mathbf{Y}_{1}-\mathbf{P}_{1})(\mathbf{Y}_{1}-\mathbf{P}_{1})^{\prime}$ and $0<\omega<1$. The control limits of the MEWMV chart are given by
\begin{equation}\label{trVi}
\textrm{E[tr}(\mathbf{V}_{i})]\pm L\sqrt{\textrm{Var[tr}(\mathbf{V}_{i})]}\,.
\end{equation}
The quantities $\textrm{E[tr}(\mathbf{V}_{i})]$ and $\textrm{Var[tr}(\mathbf{V}_{i})]$ are given in Huwang et al. (2007). 

Alfaro and Ortega (2018) changed the formulation of statistic $Z_{i}$ in (15) and proposed an alternative to the MEWMS chart for monitoring a covariance matrix which is not affected by changes in the mean. Gunaratne et al. (2017) proposed a computationally efficient algorithm based on Parallelised Monte Carlo simulation to obtain the optimal $L$ values and improved the capability of MEWMS and MEWMV charts in monitoring high-dimensional correlated quality characteristics. 
Hawkins and Maboudou-Tchao (2008) proposed yet another multivariate exponentially weighted moving covariance matrix (MEWMC) control chart similar to the MEWMS chart. To be more specific, suppose $\mathbf{A}$ is an inverse-Cholesky root matrix of $\bfmath{\Sigma}_{0}$, that is a matrix satisfying $\mathbf{A}\bfmath{\Sigma}_{0}\mathbf{A}^{\prime}=\mathbf{I}_{p}$.  Consider the transformation of $\mathbf{X}_{i}$  to $\mathbf{U}_{i}=\mathbf{A}(\mathbf{X}_{i}-\bfmath{\mu}_{0})$, which follows the distribution $N_p(\mathbf{0},\mathbf{I}_{p})$ under the in-control setting. Hawkins and Maboudou-Tchao (2008) defined $\mathbf{Z}_{i}$ as:
\begin{equation}
\mathbf{Z}_{i}=\omega \mathbf{U}_i\mathbf{U}_i^{\prime}+(1-\omega)\mathbf{Z}_{i-1}\qquad i=1,\,2,\,\dots\,,
\qquad  
\end{equation}
 where $0<\omega<1$ is the smoothing constant and $\mathbf{Z}_{0}=\mathbf{I}_{p}$. The charting statistic is then defined as:
\begin{equation}
T_{i}=\textrm{tr}(\mathbf{Z}_{i})-\textrm{log}|\mathbf{Z}_{i}|-p.
\end{equation}
Note that this statistic is exactly the same as (14) if $\mathbf{Z}_{i}$ is replaced by $\widehat{\mathbf{\Omega}}_{i}^{-1}$. The control chart signals an out-of-control situation if $T_{i}>h$, where $h$ is the upper control limit and is chosen based on the desired $\rm ARL_0$. There are two major differences between the methods in Huwang et al. (2007) and Hawkins and Maboudou-Tchao (2008). First, they used different initializations of the EWMA. Second, unlike Huwang et al. (2007), Hawkins and Maboudou-Tchao (2008) used an Alt-type likelihood ratio statistic to develop their charting statistic. These EWMA-type charts can also be modified by applying the concept of dissimilarity between two matrices. Huwang et al. (2017) used this idea and developed a multivariate exponentially weighted moving dissimilarity (MEWMD) chart that outperforms the MEWMC chart of Hawkins and Maboudou-Tchao (2008) in certain settings.  

Another variation of the methods of Huwang et al. (2007) was proposed in Memar and Niaki (2009). They employed $\ell_{1}$-norm and $\ell_{2}$-norm based distances between diagonal elements of the matrices $\mathbf{Z}_i$ in \eqref{Zi} from their expected values instead of using the trace in the control charts \eqref{trZi} proposed by Huwang et al. (2007). These modifications are also applied to MEWMV control chart. A similar attempt is done by Memar and Niaki (2011) by considering the general case of rational subgroups with $n\geq{1}$. Later, Hwang (2017) noticed that a large number of false alarms is produced when using the charts of Memar and Niaki (2009, 2011) and therefore proposed a chi-square quantile-based monitoring statistic to overcome this problem. Most recently, Ning and Li (2020) proposed employing $\ell_\frac{1}{2}$ quasinorm instead of $\ell_{1}$ and $\ell_{2}$-norms to obtain better performance than the methods proposed by Memar and Niaki (2009) and some other methods in terms of $ARL_1$. 

Recently, in a line of research similar to the case of $n\geq{p}$ discussed before, several authors have proposed methods for monitoring sparse changes in covariance matrices when $n<p$.  Yeh et al. (2012) pointed out that the method proposed by Li et al. (2013), which has been explained in Section 3, requires the subgroup size $n$ much larger than the dimensionality $p$, and thus is not applicable for individual observations. For this reason, they modified the penalty function in \eqref{PenPer1} to shrink the sample precision matrix toward the in-control one rather than to $\mathbf{0}$ and defined $\widehat{\mathbf\Omega}_{i}(\lambda)$ as:
\begin{equation}
\widehat{\mathbf{\Omega}}_{i}(\lambda)=\underset{\mathbf{\Omega}\succ\mathbf{0}}{\operatorname{argmin}}\left\lbrace {\rm tr}(\mathbf{\Omega}\mathbf{Z}_{i})-\ln|\mathbf{\Omega}|+\lambda\,\|\mathbf{\Omega}-\mathbf{I}_{p}\|_{1}\right\rbrace,
\end{equation}
where $\mathbf{Z}_{i}$ can be obtained from (15). Using this estimate together with the PLR charting statistic in (12), Yeh et al. (2012) introduced the LASSO-MEWMC (LMEWMC) chart for monitoring possible sparse changes in the covariance matrix. See also Maboudou-Tchao and Diawara (2013). 

Shen et al. (2014) observed that the performance of the proposed methods by Yeh et al. (2012) and Li et al. (2013) strongly depend on their tuning parameters, and introduced a method that overcomes this drawback. Their approach is as follows. Consider $\mathbf{Z}_{i}$ defined in (19) for any $n\geq{1}$. Let $\mathbf{C}_{i}=\mathbf{Z}_{i}-\mathbf{I}_{p}$, and $\mathbf{d}_{i}=(c_{_{i(11)}},c_{_{i(12)}},\,\cdots,\,c_{_{i(jk)}},\,\cdots,\,c_{_{i(pp)}})^\prime$, where  $c_{_{i(jk)}}$ ($j\leq k$)  is an entry in the matrix $\mathbf{C}_{i}$. 
Letting $T_{i,\,1}=\|\mathbf{d}_{i}\|_{2}$ and $T_{i,\,2}=\|\mathbf{d}_{i}\|_{\infty}$, Shen et al. (2014) introduced the MaxNorm charting statistic in the form of
\begin{equation}
{\rm max}\left\lbrace \frac{T_{i,\,1}-\textrm{E}(T_{1})}{\sqrt{\textrm{Var}(T_{1})}},\frac{T_{i,\,2}-\textrm{E}(T_{2})}{\sqrt{\textrm{Var}(T_{2})}}\right\rbrace,
\end{equation}
where $\textrm{E}(T_{1})$, $\textrm{E}(T_{2})$, $\textrm{Var}(T_{1})$, and $\textrm{Var}(T_{2})$ are estimated via a Monte Carlo simulation when $i$ is large and the process is in-control. Fan et al. (2017) proposed a control chart based on eigenvalues for monitoring covariance matrix changes, since the key features of the covariance matrix such as determinant and trace are described by these quantities. They used the eigenvalues of MEWMC defined in (19) and proposed a charting statistic similar to (22) based on these eigenvalues and showed through simulation that it outperforms the MaxNorm chart under correlation shifts.

In a recent work, Abdella et al. (2019) discussed the shortcomings of the penalized likelihood based methods in a broader fashion and argued two major flaws such methods have. First, detection performance of the penalized likelihood-based methods depends strongly on the shift patterns associated with the pre-specified tuning parameter. Second, the sparsity assumptions that are used as a building block in these methods could be violated after the data transformation step in such a way that changes in a small number of entries of the original covariance matrix may result in changes to a large number of entries in the transformed covariance matrix. Motivated by the adaptive LASSO-thresholding of Cai and Liu (2011), Abdella et al. (2019) proposed a control chart for monitoring the covariance matrix of a high dimensional process called the ALT-norm chart. Their proposed approach is not based on EWMA structure or data transformation. However, via Monte Carlo simulations, they concluded that the ALT-norm chart performs very well, in terms of $\rm{ARL}_1$, in detecting several types of shift patterns for a wide range of shift magnitudes. Kim et al. (2019) also addressed the issue of losing sparsity after data transformation and proposed a ridge ($L_2$) type penalized likelihood ratio method. This method does not rely on transformation of the data and can detect changes in the covariance matrix without a sparsity assumption when the sample size is small. 

In another attempt to remove the effect of tuning parameter, Li and Tsung (2019) integrated the powerful  test statistic of Ledoit and Wolf (2002) with an EWMA procedure for high-dimensional multivariate processes. They called their proposed MEWMV-based chart for the variability change with a large $p$ as MVP chart. Li and Tsung (2019) showed that the MVP chart is less affected by high dimensionality than the   LMEWMC, PLR, and MaxNorm charts and is more effective if the process shifts cause changes in variance components.

\section{Simultaneous monitoring of mean and covariance}
The control charts discussed in the previous sections for monitoring the covariance matrix assume that the mean vector is constant over the monitoring period. However, in practice, the mean vector may also change. Therefore, it is important to take into account the effect of a mean shift during the monitoring period, and simultaneously investigate changes in both the mean vector and the covariance matrix. There are several articles in the literature that have considered this scenario. For a bivariate process, Niaki and Memar (2009) developed a method based on the maximum of the EWMA-type charts for the transformed values of the sample mean and sample variance to jointly monitor mean vector and covariance matrix. Machado et al. (2009a) proposed two charting methods for bivariate processes: (1) the MVMAX chart that only requires the marginal sample means and variances (2) joint use of two charts based on the non-central chi-square statistic to monitor the mean vector and covariance matrix simultaneously. These ideas were extended to the general multivariate case with $p\geq2$. See Costa and Machado (2011b) and Costa and Machado (2013).

For simultaneous monitoring of the mean vector and covariance matrix, Reynolds and Cho (2006) proposed two control charts for monitoring the covariance matrix, and combined them with the standard MEWMA chart designed for monitoring the mean vector. Specifically, they let $W_{_{ijk}}=(X_{_{ijk}}-\mu_{_{0k}})/\sigma_{_{0k}},\, j=1,\cdots,\,n,\, k=1,...,p$ denote the standardized observations, and $\bfmath{\Sigma}_{_W}$ be the covariance matrix of the vector $(W_{ij1},W_{ij2},...,W_{ijp})^\prime$. Denote the respective in-control covariance matrix with $\bfmath{\Sigma}_{_{W_{0}}}$. For any $i$ and $k$, define $\overline{W}_{_{i\,.k}}=\sqrt{n}(\overline{X}_{i\,.k}-\mu_{0k})/\sigma_{0k}$, where $\overline{X}_{i\,.k}=n^{-1}\sum_{j=1}^{n}{X_{_{ijk}}}$ is the sample mean for the $k$th variable at the sampling epoch $i$.
The EWMA statistics of the standardized sample mean and the squared standardized deviations from target for the $k$th variable are defined by
\begin{align}
E_{_{ik}}^{W}&=(1-\gamma)E_{_{(i-1)k}}^{W}+\gamma \overline{W}_{_{i\,.k}}\,,\\
E_{_{ik}}^{W^2}&=(1-\gamma)E_{_{(i-1)k}}^{W^2}+ \frac{\gamma}{n}\sum_{j=1}^{n}{W_{_{ijk}}^2},
\end{align}
respectively, where $E_{_{0k}}^W=0$ and $E_{_{0k}}^{W^2}=1$ are the initial values, and $0<\gamma\leq1$ is the tuning parameter. Assuming $\bfmath{\Sigma}_{_0}$ is given, the MEWMA control chart for monitoring the mean vector is given by the quadratic form 
\begin{equation}
M_{_i}^{W}=b_{_\infty}^{-1}(E_{_{i1}}^{W},\,E_{_{i2}}^{W},\,\cdots,\,E_{_{ip}}^{W})\,\bfmath{\Sigma}_{_{W_{_0}}}^{-1}\,(E_{_{i1}}^{W},\,E_{_{i2}}^{W},\,\cdots,\,E_{_{ip}}^{W})^{\prime},
\end{equation}
where $b_{_\infty}=\gamma/(2-\gamma)$. In addition, the two MEWMA-type statistics of Reynolds and Cho (2006) for monitoring $\bfmath{\Sigma}$ are 
\begin{align*}
M_{_{1i}}^{W^2}&=\frac{n}{2\,b_{_\infty}}\,\left(E_{_{i1}}^{W^2}-1,\,E_{_{i2}}^{W^2}-1,\,\cdots,\,E_{_{ip}}^{W^2}-1\right)\,\left(\bfmath{\Sigma}_{_{W_{0}}}^{(2)}\right)^{-1}\,\left(E_{_{i1}}^{W^2}-1,\,E_{_{i2}}^{W^2}-1,\,\dots,\,E_{_{ip}}^{W^2}-1\right)^{\prime},\\
M_{_{2i}}^{W^2}&=\frac{n}{2\,b_{_\infty}}\,\left(E_{_{i1}}^{W^2},\,E_{_{i2}}^{W^2},\,\cdots,\,E_{_{ip}}^{W^2}\right)\,\left(\bfmath{\Sigma}_{_{W_{0}}}^{(2)}\right)^{-1}\,\left(E_{_{i1}}^{W^2},\,E_{_{i2}}^{W^2},\cdots,\,E_{_{ip}}^{W^2}\right)^{\prime},
\end{align*}
where $\left(\bfmath{\Sigma}_{_{W_{0}}}^{(2)}\right)^{-1}$ represents the inverse of a matrix whose entries are the square of the corresponding entries of the matrix $\bfmath{\Sigma}_{_{W_{0}}}$. The chart $M_{1i}^{W^2}$ is implemented with a UCL, while $M_{2i}^{W^2}$ is used with both an LCL and a UCL and is more effective in detecting decreases in variability. Reynolds and Cho (2006) studied the performance of the aforementioned MEWMA control charts and compared them to some other existing EWMA-type and Shewhart-type control charts in the literature. See also Reynolds and Stoumbos (2008, 2010), Reynolds and Kim (2007), and Reynolds and Cho (2011). 
It is worthwhile to mention that, despite their effectiveness, the hybrid approaches similar to  those in Reynolds and Cho (2006) suffer from not providing detailed diagnostic information when a process is out-of-control. For example, in simultaneous application of the Hotelling $T^{2}$ (or MEWMA) and $|\,\mathbf{S}\,|$ control charts, the former may react to mean shifts, dispersion changes, and changes of correlations, which will mask the nature of shift. This is mainly because those charts assume that the covariance matrix (or its estimate) remains unchanged during the monitoring process. Zhang and Chang (2008) proposed a combined charting scheme for monitoring individual observations which differentiates mean shifts from shifts in covariance matrix. Their proposal is composed of two control charts. The first control chart is a dynamic MEWMA (DMEWMA) that sequentially updates the estimate of the covariance matrix as new data become available. Hence, this chart is sensitive to mean shifts only. The second chart is a multivariate exponentially weighted moving deviation (MEWMD) control chart which monitors the difference between the current observation and the current process mean estimated by a moving average of a certain number of the most recent observations. Note that this MEWMD chart detects shifts in covariance matrix only. 

Instead of implementing two separate control charts, Zhang et al. (2010) proposed a single control chart based on integrating an EWMA procedure with the generalized LRT. This new chart is similar to the MEWMC chart proposed by Hawkins and Maboudou-Tchao (2008) but is designed to monitor the process mean vector and covariance matrix simultaneously. This control chart can also be used for the case of individual observations. Wang et al. (2014) used the advantage of the penalized likelihood estimation in producing sparse estimates when only a small set of the mean or variance/covariance components contributes to changes in the process. They proposed two control charts based on the penalized likelihood estimate for multivariate processes in which only individual observations are available. Their idea extends the previously discussed work of Li et al. (2013) and Yeh et al. (2012) to simultaneously estimate the process mean vector and covariance matrix via the penalized likelihood method and construct the relevant charting statistic.

A common difficulty in multivariate statistical process monitoring is obtaining good estimates of process parameters and this often requires a large Phase I dataset. Zamba and Hawkins (2009) developed a multivariate change-point analysis through generalized likelihood ratio statistic applied sequentially for detecting changes in the mean and/or covariance matrix as well as their locations. Their unknown-parameter change-point formulation has the benefit of not needing a large Phase I dataset. Maboudou-Tchao and Hawkins (2011) considered an unknown-parameter self-starting formulation, which allows the Phase II monitoring to start without needing a large Phase I dataset, as they do not require parameters estimation before beginning Phase II. Maboudou-Tchao and Hawkins (2011) extended the existing multivariate self-starting charts for the mean vector and covariance matrix to a combination control scheme for the joint monitoring of mean and covariance. The basic idea of the method is to transform the original process readings into a stream of mutually independent and identically distributed data for which the in-control distribution is known and can then be used in known-parameter control charts. Self-starting control charts use successive observations for recalculating the parameter estimates each time a new sample is taken while at the same time checking for the process stability.  

\section{Other Relevant Topics}
In this section, we briefly discuss some selected work on other topics relevant to covariance matrix monitoring. This includes diagnostic approaches, non-parametric methods for Phase II, and Phase I analysis among others. 

\subsection{Diagnosis of out-of-control signals}
Once a control chart has signalled an alarm, it is important to identify the cause by performing diagnostic analysis on the magnitude of shift and the time of change. Since a covariance matrix contains many unknown parameters, the diagnostic step in a covariance matrix monitoring is a much more involved task than that for a mean vector. The bivariate case is relatively simple and detection methods such as those of Costa and Machado (2008) and Niaki and Memar (2009) can be used to find the out-of-control quality characteristics. However, in higher dimensional settings, finding a characteristic(s) responsible for a signal is not an easy task. Sullivan et al. (2007) extended the step-down diagnostic procedure used in mean vector monitoring to monitoring a covariance matrix. Let, collectively, $\bfmath{\theta}$ consist of the elements in the mean vector and the unique elements of the covariance matrix. For example, for a bivariate normal distribution with in-control correlation coefficient $\rho_{_0}$, the vector $\bfmath{\theta}$ under in-control situation can be defined as $\bfmath{\theta}_{0}=({\mu}_{_{0\,1}},{\mu}_{_{0\,2}},{\sigma}_{_{0\,1}},\rho_{_0}, {\sigma}_{_{0\,2}})$. Let  $\bfmath{\theta}_{0}$ ($\bfmath{\theta}_{1}$) be the parameter vector of the observations before (after) the change point and $\widehat{\bfmath{\theta}}_{0}$ ($\widehat{\bfmath{\theta}}_{1}$) be the corresponding maximum likelihood estimate. Define the difference vector $\bfmath{\delta}=(\bfmath{\theta}_{0}-\bfmath{\theta}_{1})$ with $k_0$ elements and $\widehat{\bfmath{\delta}}=(\widehat{\bfmath{\theta}}_{0}-\widehat{\bfmath{\theta}}_{1})$ as its maximum likelihood estimate. As the maximum-likelihood parameter estimators are
asymptotically multivariate normal, the statistic  $\widehat{\bfmath{\delta}}^{\prime}\widehat{\bfmath{\Sigma}}^{-1}_{\widehat{\bfmath{\delta}}}\widehat{\bfmath{\delta}}$ asymptotically has the chi-square distribution with $k_0$ degrees of freedom,
where ${\bfmath{\Sigma}}_{\hat{\bfmath{\delta}}}$ is the covariance matrix of $\hat{\bfmath{\delta}}$ and $\hat{\bfmath{\Sigma}}_{\hat{\bfmath{\delta}}}$ is its estimator. By using this property, the step-down approach proposed by Sullivan et al. (2007) partitions the parameter set into two subsets, in which one subset contains those parameters whose estimates are statistically different on the two sides of change point and the other subset includes parameters for which their estimates are not statistically different. 
This procedure then searches for the largest subset that have the feature of no evidence of a change and then selects the remaining parameters as shifted parameters. This method is also applicable for non-normal multivariate data. Mingoti and Pinto (2018) showed through simulation that with few exceptions, the method of Sullivan et al. (2007) has better power to detect out-of-control situations than that of both VMAX and VMIX control charts under the bivariate normal assumption. Cheng and Cheng (2008) formulated identifying the source of shifts in the covariance matrix of a multivariate process as a classification problem and proposed two classifiers based on neural networks (NN) and support vector machines (SVM) to identify which variable is responsible for the shift. See also Salehi et al. (2012) and Cheng and Lee (2016) in this regard. Mason et al. (2010) proposed a decomposition of Wilks' ratio statistic, introduced in Section 4, to identify the process variables contributing to a signal. Hung and Chen (2012) applied Bartlett's decomposition and Cholesky's decomposition to classify changes in a covariance matrix and to determine the source of fault without a large sample requirement. Zou et al. (2011) considered the sparsity feature in high-dimensional processes and argued that a fault diagnosis problem is similar to the variable or model selection problem in this setting. Suppose that the true fault isolation model contains all the indices corresponding to the parameters that have really shifted and only those indices. If $A=\lbrace 1,\,\cdots,\,\ell\rbrace$, is the full set of the indices of the parameters and $B$ is a candidate subset of $A$ that identifies all the indices corresponding to the changed parameters, the objective is to select an optimal $B\subset A$ so that $B$ is as close as possible to the true fault isolation model. For this purpose, Zou et al. (2011) reduced the fault isolation to the two-sample variable selection problem and provided a unified diagnosis framework based on Bayesian information criterion (BIC). Zou et al. (2011) then proposed combining BIC with the adaptive LASSO variable selection method (LEB method) to obtain a LASSO-based diagnostic procedure which includes diagnosis in both the mean vector and the covariance matrix under the sparsity assumption of parameter changes. Wang et al. (2020) developed a method using Bayesian variable selection method to diagnose shifts only in the covariance matrix and used a Gibbs sampling procedure to overcome the computational problems of their proposed Bayesian hierarchical model. They compared this to LEB through simulations and concluded that their method has better overall performance, especially under shifts in correlations.

Some work focused on change point estimation as an important part of diagnosing control chart signals. Dogu and  Deveci-Kocakoc (2011) considered change point estimation when Alt's control charts based on generalized variance (introduced in Section 3) are used for monitoring a covariance matrix and derived the maximum likelihood estimation (MLE) of a change point in the covariance matrix only. Similarly, Dogu and  Deveci-Kocakoc (2013) proposed a change point estimation procedure based on MLE when the $T^2$ and the generalized variance control charts are used to simultaneously monitor the mean vector and covariance matrix. Moreover, Dogu (2015) considered change point detection using MLE when some EWMA-based control charts are used simultaneously to monitor the mean vector and covariance matrix.

\subsection{Phase II monitoring under non-normality of data}
Multivariate control charts are generally constructed assuming normality. It is generally well understood that these procedures lose their efficiency when the process distribution deviates from multivariate normality. Therefore, robust control charts are preferable and distribution-free or nonparametric control charts are designed to achieve this purpose. As mentioned by Chakraborti et al. (2001), while the term "distribution-free" seems to be a better description of what we expect these charts to accomplish, "nonparametric" is the term most often used, and hence, both distribution-free and nonparametric describe the same charts. Riaz and Does (2008) used the sample Gini mean differences based matrix, denoted by $\mathbf{G}$, as an estimate of the covariance matrix and proposed a robust control chart for Phase II monitoring of the process variability in a bivariate process when departure from bivariate normality is suspected. For a bivariate normal process with variables $X_1$ and $X_2$, the matrix $\mathbf{G}$ is defined as
\begin{equation}
 \mathbf{G}=\begin{bmatrix} G^{2}_{1} & G_{12} \\ G_{21} & G^{2}_{2} \end{bmatrix},
\end{equation}
 where 
 \begin{align*}
 G_{1}=(2\sqrt{\pi}){\rm cov}(X_1,F(X_1)),\quad G_{2}=(2\sqrt{\pi}){\rm cov}(X_2,F(X_2)),\\
  G_{12}=(2\sqrt{\pi}){\rm cov}(X_1,F(X_2))\,G_{2},\quad G_{21}=(2\sqrt{\pi}){\rm cov}(X_2,F(X_1))\,G_{1}
 \end{align*}
and where $F(\cdot)$ denotes the cumulative distribution function and ${\rm cov}(X_1,X_2)$ is the covariance between the two variables. The generalized Gini mean differences can be then defined as $|\mathbf{G}|=G^{2}_{1}.G^{2}_{2}-G_{12}.G_{21}$. Riaz and Does (2008) developed the $\mathbf{G}$ control chart based on distributional properties of $|\mathbf{G}|^\frac{1}{2}$ assuming that both variable $X_1$ and $X_2$ are normally distributed and examined the departures from normality for the proposed $|\, \mathbf{G}\,|$ chart. But as pointed out by Saghir (2015), the design structures of both $|\,\mathbf{S}\,|$ and $|\, \mathbf{G}\,| $ charts are based on multivariate normal distribution. Therefore, Saghir (2015) considered a variety of non-normal distributions and proposed asymmetrical probability limits for $|\,\mathbf{S}\,|$ and $|\, \mathbf{G}\,| $ charts for each underlying distribution in Phase II monitoring of bivariate non-normal processes. For a $|\, \mathbf{G}\,| $ chart, the probability upper and lower control limits are defined as 
 \begin{align*}
 LCL=\frac{|\bfmath{\Sigma}_{0}|^\frac{1}{2}}{2(n-1)}B_{(1-\alpha/2),n},\quad UCL=\frac{|\bfmath{\Sigma}_{0}|^\frac{1}{2}}{2(n-1)}B_{\alpha/2,n},
 \end{align*} 
where $B_{\alpha/2,n}$ and $B_{(1-\alpha/2),n}$ are the quantile points of the distribution of $B$, which is a random variable that defines the relationship between $|\mathbf{G}|^\frac{1}{2}$ and the square root of the population generalized variance $|\bfmath{\Sigma}|^\frac{1}{2}$ defined as $B=2(n-1)|\mathbf{G}|^\frac{1}{2}/|\bfmath{\Sigma}|^\frac{1}{2}$. Saghir (2015) suggested good choices of quantiles points according to the parent populations to achieve a better performance. Zhou et al. (2015) proposed a distribution-free procedure by integrating a multivariate two-sample goodness-of-fit (GOF) test based on  minimal spanning tree (MST) and a change-point model. It is shown to be efficient especially in detecting moderate to large process shifts through simulation. Zhang et al. (2016) highlighted a drawback of nonparametric control charts in the sense that although they robustly detect changes across different types of data distributions, they are not totally distribution-free, that is, the specified in-control run-length distribution depends on the exact in-control distribution. The other limitation is that the conventional nonparametric charts require a sufficiently large in-control samples and cannot perform well with limited in-control samples. Zhang et al. (2016) proposed a distribution-free multivariate control chart based on a multivariate goodness of fitness (GoF) test to detect general distributional changes (including changes in the mean vector, the covariance matrix, and the distribution shapes). Their chart employs data-dependent control limits and is also applicable for high-dimensional observations. They showed that it has satisfactory performance even with an unknown IC distribution or limited reference samples. 
Osei-aning et al. (2017) suggested use of the maximum of dispersion estimates such as the sample standard deviation, interquartile range, average absolute deviation from median and the median absolute deviation and developed Phase II control charts based on these statistics for normal and non-normal bivariate processes. 
Mostajeran et al. (2017) developed non-parametric bootstrap versions of some existing multivariate control charts for monitoring a covariance matrix for the situations where the normality assumption of distribution might be violated and collecting a large Phase I sample is not possible. Haq and Khoo (2018) proposed a non-parametric MEWMA sign chart which is robust to the violation of normality in the simultaneous monitoring of mean and covariance matrix. At sampling epoch $i$, define the random variable $A_{i}=\sum_{j=1}^{n}{I_{ij}}$, where $I_{ij}$ is an indicator variable with value 1 if $({\mathbf{X}_{i\,j}}-\bfmath{\mu}_{_0})'\bfmath{\Sigma_{0}^{-1}}({\mathbf{X}_{i\,j}}-\bfmath{\mu}_{_0})>p$ and zero, otherwise. Then, $A_{i}$ is a binomial random variable with parameters $n$ (number of independent Bernoulli trials) and probability $\pi_0$ (probability of a single success which depends on the underlying probability distribution of process) under the in-control situation. Haq and Khoo (2018) proposed a non-parametric EWMA-type control chart for monitoring $B_i=\textrm{sin}^{-1}(\sqrt{A_{i}/n})$, which can also be used to simultaneously monitor changes in the process mean and covariance matrix.
 Liang et al. (2019) applied the spatial sign covariance matrix and maximum norm to the EWMA scheme to propose a robust multivariate control chart similar to the MaxNorm chart proposed by Shen et al. (2014) for normal processes (Section 4). They propose transforming the original vector of quality characteristics $\mathbf{X}_i$, to the unit vector $\mathbf{U}_i$ through
 \begin{align*}
\mathbf{U}_{i}=\frac{\mathbf{A}_{0}(\mathbf{X}_{i}-\bfmath{\theta}_{0})}{||\mathbf{A}_{0}(\mathbf{X}_{i}-\bfmath{\theta}_{0})||_{2}}\,,
 \end{align*}
where $\bfmath{\theta}_{0}$ is a $p\times{1}$ vector called the multivariate median vector and $\mathbf{A}_0$ is a $p\times{p}$ upper triangular positive-definite transformation matrix. Both $\bfmath{\theta}_{0}$ and $\mathbf{A}_0$ can be estimated based on sample equations and from Phase I historical data. By using $\mathbf{U}_{i}$, the corresponding EWMA statistic $Z_i$ can be given from equation (19) with the initial vector $\mathbf{I}_{p}/p$. Then, one can define $\mathbf{C}_{i}=\mathbf{Z}_{i}-\mathbf{I}_{p}/p$, and $\mathbf{d}_{i}$, $T_{i,\,1}$, $T_{i,\,2}$, and a charting statistic similar to those in Shen et al (2014). Liang et al. (2019) showed that the resulting chart, called SMaxNorm chart, is distribution-free over the family of elliptical distributions and is more sensitive than PLR, MaxNorm, and LMEWMC (explained in Sections 3 and 4), for various covariance matrix shifts under heavy-tailed distributions. Koutras and Sofikitou (2020) proposed two semiparametric control charts by exploiting the order statistics for simultaneously identifying shifts in the mean and covariance of a bivariate process. Declaring an out-of-control in both proposed charts is based on checking both the location of a specific order statistic like the median and the number of specific observations lying between the upper and the lower control limits. Although the charts in Koutras and Sofikitou (2020) formally depend on the bivariate dependence structure, the effect of this dependence structure on them is negligible, and thus the authors suggest these charts to be categorized as nonparametric.

\subsection{Phase I analysis}
As previously mentioned, Phase I data are used to gain process knowledge, assess process stability, and estimate process parameters. Phase I is difficult and critical in process monitoring since the additional variability introduced from the estimation step affects the Phase II control chart's performance. The success of Phase I methods depends on their ability to distinguish the outlying observations correctly and provide reliable estimates of the process parameters, which is related to the correct specification of the underlying in-control model (Jones-Farmer et al. 2014). Two widely used multivariate high breakdown robust estimators for multivariate location and scatter are minimum volume ellipsoid (MVE) and minimum covariance determinant (MCD). While the MVE finds the ellipsoid with minimum volume that covers a subset of at least half of the observations, the MCD's objective is to find a fixed-size subset of the data whose covariance matrix has the lowest determinant. Variyath and Vattathoor (2014) used two robust estimators including the re-weighted minimum covariance determinant (RMCD) and the re-weighted minimum volume ellipsoid (RMVE) and proposed RMCD/RMVE- based MEWMS/MEWMV control charts for Phase I monitoring of a covariance matrix in multivariate processes with individual observations. See also Vargas and Lagos (2007) in this regard.
Li et al. (2014) applied the concept of data depth and the Mann-Whitney test and proposed a  nonparametric multivariate Phase I control chart for detecting shifts in both the mean vector and the covariance matrix of processes with individual observations and estimated the change point in Phase I based on the MLE. Saghir et al. (2016) proposed control limits for Phase I $|\mathbf{S}|$ and $|\mathbf{G}|$ charts for a bivariate normal process based on the false alarm probability (FAP). Saghir et al. (2017) extended this work to bivariate non-normal distributions and provided the necessary constants, which are required in the construction and implementation of these charts, through an extensive simulation. Abdella et al. (2020) made the assumption of a sparse covariance matrix in high-dimensional multivariate normal processes and proposed using the adaptive thresholding LASSO rule for estimating the unknown covariance matrix and constructing control chart in Phase I analysis of a covariance matrix. Recently, using a two-sample test for high-dimensional covariance matrices, which compares the difference in the leading eigenvalues for comparing two high-dimensional covariance matrices, Fan et al. (2020) developed a non-parametric Phase I control chart which showed good performance specially when the shift in covariance matrix is weak and sparse.
Recall that in Phase I, $N$ samples are available based on historical data and  suppose that $1\leq{\tau}\leq{N-1}$ is the potential change point and ${\tau}^*$ is the unknown change point in Phase I. For each $\tau$, let the sample covariance matrices before and after the potential change point be denoted by  $\widehat{\bfmath{\Sigma}}_{0, \tau}$ and $\widehat{\bfmath{\Sigma}}_{1, \tau}$, respectively. These estimates are calculated by using Phase I observations before and after the sampling epoch $\tau$. For $\tau=1,\,\dots,\,N-1$, define
\begin{equation}
\mathbf{D}_{\tau}=\widehat{\bfmath{\Sigma}}_{0, \tau}-\widehat{\bfmath{\Sigma}}_{1, \tau}\,,
\end{equation}
and let
\begin{equation}
T(\tau)=\,{\textrm{max}}\,\{|\eta_{\,\textrm{max}}^{\,R}(\mathbf{D}_\tau)|,|\eta_{\,\textrm{max}}^{\,R}(\mathbf{-D}_\tau)|\}\, ,
\end{equation}
where $\eta_{\,\textrm{max}}^{\,R}(\mathbf{D}_\tau)$ is the R-sparse-leading eigenvalue of the differential matrix $\mathbf{D}_{\tau}$. For any symmetric matrix $\mathbf{A}$, the R-sparse-leading eigenvalue can be obtained by solving the following constrained optimization problem
\begin{equation}
\eta_{\,\textrm{max}}^{\,R}(\mathbf{A})=\underset{||\bfmath{\nu}||_{_2}=1,\, ||\bfmath{\nu}||_{_0}\leq{R}}{{\rm max}} {\rm tr}(\mathbf{A}(\bfmath{\nu}\bfmath{\nu^{'}}))\, ,
\end{equation}
where $\bfmath{\nu}$ is a $p\times{1}$ vector, $||\cdot||_{_0}$ denotes the $\ell_0$-norm, and the value of $R$ controls sparsity of $\bfmath{\nu}$. Since when $\tau$ is relatively small, the sample covariance estimate $\widehat{\bfmath{\Sigma}}_{0, \tau}$ could be degenerated,  Fan et al. (2020) considered a minimum window length of $w$ observations and proposed a Phase I control chart which signals whenever the following statistic exceeds a threshold
\begin{equation}
\underset{w\leq{\tau}\leq{N-w}}{\textrm{max}} T(\tau)\, .
\end{equation}
Once this chart triggers an alarm, then the corresponding $\tau^*$ can be also estimated through
\begin{equation}
\hat{\tau}^*=\underset{w\leq{\tau}\leq{N-w}}{\textrm{argmax}} T(\tau)\, .
\end{equation}

For convenience of access, a summary of the methods for monitoring the covariance matrix is provided in Table 1 and a  yearly-sorted classification of all methods related to this topic is presented in Table 2.

\begin{sidewaystable}[]
\renewcommand{\arraystretch}{1.3}
\tiny
\centering
\caption{A summary of methods for monitoring covariance matrix}
\begin{tabular}{|p{3cm}|p{8.1cm}|p{4.5cm}|p{1.5cm}|p{1.4cm}|p{1.95cm}|} 
 \hline
 & & &&&\\
 \centering{Problem}&	\centering{Idea/Chart} &	\centering{Reference}&	\centering{Process Type} &	 \centering{Sparsity Assumption}&	\centering\arraybackslash Only Individual Observations  \\
 & & &&&  \\
  \hline
 & & & & & \\
 & Generalized variance $|\,\mathbf{S}\,|$ chart& Alt (1985)& Multivariate  & \centering\arraybackslash $\times$ & \centering\arraybackslash $\times$ \\
&	Synthetic control chart (combining $|\mathbf{S}|$ and CRL charts)& Ghute \& Shirke (2008)&		Multivariate  & \centering\arraybackslash $\times$ & \centering\arraybackslash $\times$ \\
 &  Group runs $|\mathbf{S}|$ chart & Gadre (2014)& Multivariate&   \centering\arraybackslash $\times$  &\centering\arraybackslash $\times$  \\
 &  Synthetic $|\mathbf{S}|$ control charts &Lee \& Khoo (2015,2017a,b)	& Multivariate  & \centering\arraybackslash $\times$  &\centering\arraybackslash $\times$  \\
  & VMAX statistic (based on the marginal sample variances) & Costa \& Machado (2009)&Multivariate &\centering\arraybackslash $\times$ & \centering\arraybackslash $\times$ \\
   & A variant of VMAX & Costa \& Machado (2008)& Bivariate  & \centering\arraybackslash $\times$ & \centering\arraybackslash $\times$ \\
 & A variant of VMAX  (VMAX Group Runs) &Gadre \& Kakade (2018) & Bivariate  & \centering\arraybackslash $\times$ &\centering\arraybackslash $\times$  \\
 &A synthetic chart based on the VMAX & Machado et al. (2009b) &Bivariate    &\centering\arraybackslash $\times$  & \centering\arraybackslash $\times$ \\
 Covariance monitoring& Use of 3 attribute control charts & Machado et al. (2018)	 & Bivariate  &\centering\arraybackslash $\times$ & \centering\arraybackslash $\times$ \\
 when  $n\geq{p}$ (Section 3)&An statistic based on the sample variances & Costa \& Neto (2017) &2 \& 3-variate   & \centering\arraybackslash $\times$ & \centering\arraybackslash $\times$ \\
 & VMIX statistic (based on  stacked sample variance)& Quinino et al. (2012)& Bivariate & \centering\arraybackslash$\times$ & \centering\arraybackslash $\times$ \\
 & Artificial Neural Network & Cheng \& Cheng (2011) &Bivariate  & \centering\arraybackslash $\times$ & \centering\arraybackslash $\times$ \\
 &RMAX statistic (based on the marginal sample ranges)& Costa \& Machado (2011a)& Multivariate & \centering\arraybackslash $\times$  &\centering\arraybackslash $\times$  \\
 &LRT statistic of a one-sided test (increase in variances) & Yen \& Shiau (2010)&Multivariate   &\centering\arraybackslash $\times$  & \centering\arraybackslash $\times$ \\
 & LRT statistic of a one-sided test (decrease in variances)& Yen et al. (2012) & Multivariate  & \centering\arraybackslash $\times$ & \centering\arraybackslash $\times$ \\
 & Penalized Likelihood Ratio (PLR) statistic	& Li et al. (2013)	&Multivariate & \centering\arraybackslash \checkmark &\centering\arraybackslash $\times$ \\
  & MCUSUM control chart & Healy (1987)	&Multivariate & \centering\arraybackslash $\times$ &\centering\arraybackslash $\times$ \\
  &  CUSUM charts and Shewhart-type control charts	& Surtihadi et al. (2004) 	&Multivariate & \centering\arraybackslash $\times$ &\centering\arraybackslash $\times$ \\
    &  EWMA charts based on the VMAX	&  Machado \& Costa (2008) 	&Bivariate & \centering\arraybackslash $\times$ &\centering\arraybackslash $\times$ \\
    &  EWMA-type control charts &   Osei-Aning \& Abbasi (2020) 	&Bivariate & \centering\arraybackslash $\times$ &\centering\arraybackslash $\times$ \\
     &  Multivariate Mixed EWMA-CUSUM (MMECD) control chart &    Riaz et al. (2019) 	&Multivariate & \centering\arraybackslash $\times$ &\centering\arraybackslash $\times$ \\
  & & & & &\\

 \hline
 & & & & &\\
&  Shewhart-type charts based on  vector variance (VV)& Djauhari et al. (2008) &Multivariate & \centering\arraybackslash $\times$ & \centering\arraybackslash $\times$ \\ 
 & Wilks' statistic& Mason et al. (2009) &  Multivariate& \centering\arraybackslash $\times$ & \centering\arraybackslash $\times$ \\
  & Shewhart-type chart by modification of PLR chart & Maboudou-Tchao \& Agboto (2013)& Multivariate  & \centering\arraybackslash \checkmark & \centering\arraybackslash $\times$ \\
 & Two exponentially weighted chart (MEWMS and MEWMV)& Huwang et al. (2007)&Multivariate & \centering\arraybackslash $\times$ & \centering\arraybackslash \checkmark \\
  &A modification to MEWMS chart& Alfaro \& Ortega (2018)& Multivariate & \centering\arraybackslash $\times$ & \centering\arraybackslash \checkmark \\
 &An exponentially weighted chart (MEWMC) & Hawkins \& Maboudou-Tchao (2008)& Multivariate & \centering\arraybackslash $\times$ & \centering\arraybackslash \checkmark \\
 & An exponentially weighted chart (MEWMD)&Huwang et al. (2017)& Multivariate  &  \centering\arraybackslash $\times$ & \centering\arraybackslash $\times$ \\
Covariance monitoring &Modification of MEWMS and MEWMV charts& Memar \& Niaki (2009) &Multivariate & \centering\arraybackslash $\times$ & \centering\arraybackslash \checkmark \\
  when $n<p$ (Section 4)& Modification of MEWMS and MEWMV charts& Memar \& Niaki (2011) &Multivariate & \centering\arraybackslash $\times$ & \centering\arraybackslash $\times$ \\
 & Chi-square quantile-based monitoring statistic & Hwang (2017)&Multivariate & \centering\arraybackslash $\times$  & \centering\arraybackslash \checkmark \\
 &LASSO-MEWMC (LMEWMC) chart& Yeh et al. (2012)&Multivariate& \centering\arraybackslash \checkmark & \centering\arraybackslash \checkmark \\

 &MaxNorm charting statistic	& Shen et al. (2014)&Multivariate & \centering\arraybackslash \checkmark & \centering\arraybackslash $\times$ \\
  & A chart based on the eigenvalues of MEWMC& Fan et al. (2017) & Multivariate & \centering\arraybackslash $\times$ & \centering\arraybackslash $\times$ \\
 & Adaptive LASSO-Thresholding (ALT) Norm chart& Abdella et al. (2019)& Multivariate & \centering\arraybackslash \checkmark & \centering\arraybackslash $\times$ \\
 & Ridge ($L_2$) type penalized likelihood ratio method& Kim et al. (2019)& Multivariate& \centering\arraybackslash $\times$  & \centering\arraybackslash $\times$ \\
 &Integrating a high-dimensional two-sample test with EWMA procedure& Li \& Tsung (2019)&  Multivariate & \centering\arraybackslash $\times$ & \centering\arraybackslash $\times$ \\
 & & & & &\\

 \hline
 & & & & &\\
  & Based on the maximum of the usual EWMA charts& Niaki \& Memar (2009)&  Bivariate& \centering\arraybackslash $\times$ & \centering\arraybackslash $\times$ \\
 & Two control charts (MVMAX and non-central chi-square statistic) & Machado et al. (2009a)&  Bivariate& \centering\arraybackslash $\times$  & \centering\arraybackslash $\times$ \\
 & Extensions of  Machado et al. (2009a) chart to multivariate case& Costa \& Machado (2011b , 2013)&  Multivariate  & \centering\arraybackslash $\times$ & \centering\arraybackslash $\times$ \\
 & Combination of MEWMA-type control charts for covariance and mean& Reynolds \& Cho (2006) &  Multivariate  & \centering\arraybackslash $\times$ &\centering\arraybackslash $\times$  \\
 Simultaneous monitoring & Modifications to Reynolds \& Cho (2006)	& Reynolds \& Stoumbos (2008, 2010) &Multivariate & \centering\arraybackslash $\times$ & \centering\arraybackslash $\times$ \\
 of mean and covariance& Modifications to Reynolds \& Cho (2006) & Reynolds \& Kim (2007)&  Multivariate & \centering\arraybackslash $\times$ & \centering\arraybackslash $\times$ \\
(Section 5)& Modifications to Reynolds \& Cho (2006)& Reynolds \& Cho (2011)&  Multivariate & \centering\arraybackslash $\times$ & \centering\arraybackslash $\times$ \\
 &Combination of two charts (DWEMA and MEWMD) & Zhang \& Chang (2008)& Multivariate& \centering\arraybackslash $\times$ & \centering\arraybackslash \checkmark \\
 & Integrating EWMA procedure with the generalized LRT	& Zhang et al. (2010) &  Multivariate & \centering\arraybackslash $\times$ & \centering\arraybackslash $\times$ \\
 & Two control charts based on the penalized likelihood estimate & Wang et al. (2014) &  Multivariate & \centering\arraybackslash \checkmark & \centering\arraybackslash \checkmark \\
 &Multivariate change-point analysis (generalized likelihood ratio statistic)& Zamba \& Hawkins (2009)&  Multivariate& \centering\arraybackslash $\times$ & \centering\arraybackslash \checkmark\\
 &Self-starting control schemes for the mean and covariance& Maboudou-Tchao and Hawkins (2011)&  Multivariate& \centering\arraybackslash $\times$ & \centering\arraybackslash \checkmark\\
 & & & & &\\
 \hline
\end{tabular}
\end{sidewaystable}

\begin{sidewaystable}[]
\renewcommand{\arraystretch}{1.2}
\tiny
\centering
\caption{A classification of the MSPM methods developed for covariance matrix}
\begin{tabular}{|p{4.05cm}|>{\centering}p{2.5cm}|>{\centering}p{1.4cm}|>{\centering}p{1.9cm}||p{3.4cm}|>{\centering}p{2.1cm}|>{\centering}p{2.3cm}|p{3.4cm}|}\tabularnewline
 \hline
 & & &&&&&\\
 \centering{Reference}&\centering{Purpose} & \centering{Distribution} &	  \centering{Chart Type}&	\centering{Reference}&	\centering{Purpose} & \centering{Distribution} &	\centering{Chart Type} \tabularnewline  
 & & &&&&&\\ 
 \hline
 & & &&&&&\\
Alt (1985)&Phase II&Normal&Shewhart& Gadre (2014) &Phase II&Normal& \centering\arraybackslash Shewhart (group runs)\\
 Healy (1987)& Phase II&Normal&CUSUM& Li et al. (2014)&Phase I&(Non) Normal&\centering\arraybackslash Non-parametric\\
 Surtihadi et al. (2004)&Phase II&Normal&CUSUM& Shen et al. (2014)&Phase II&Normal&\centering\arraybackslash EWMA\\
 Reynolds \& Cho (2006)&Phase II&Normal&EWMA& Variyath \& Vattathoor (2014)&Phase I&Normal&\centering\arraybackslash EWMA\\ 
 Chang \& Zhang (2007)&Phase II&Normal&EWMA& Wang et al. (2014)&Phase II&Normal&\centering\arraybackslash EWMA\\
 Huwang et al. (2007)&Phase II&Normal&EWMA& Dogu (2015)&Change point&Normal&\centering\arraybackslash EWMA\\
 Reynolds \& Kim (2007)&Phase II&Normal&EWMA& Lee \& Khoo (2015)&Phase II&Normal&\centering\arraybackslash Shewhart (adaptive)\\
 Sullivan et al. (2007)&Diagnosis&(Non)Normal&-& Saghir (2015)&Phase II&(Non) Normal&\centering\arraybackslash Shewhart\\
 Vargas \& Lagos (2007)&Phase I&Normal&Shewhart& Zhou et al. (2015)&Phase II&(Non) Normal&\centering\arraybackslash Non-parametric\\ 
 Cheng \& Cheng (2008)&Diagnosis&Normal&-& Cheng \& Lee (2016)&Diagnosis&Normal&\centering\arraybackslash -\\
Costa \& Machado (2008)&Phase II&Normal &Shewhart& Djauhari et al. (2016)&Phase II&Normal& \centering\arraybackslash Shewhart\\ 
 Djauhari et al. (2008)&Phase II&Normal&Shewhart& Ramos et al. (2016)&Misleading signals&Normal&\centering\arraybackslash EWMA/Shewhart\\  
 Ghute \& Shirke (2008)&Phase II&Normal&Shewhart (syn)& Saghir et al. (2016)&Phase I&Normal&\centering\arraybackslash Shewhart\\
 Hawkins \& Maboudou-Tchao (2008)&Phase II&Normal&EWMA& Zhang et al. (2016)&Phase II&(Non) Normal&\centering\arraybackslash Non-parametric\\
 Machado \& Costa (2008)&Phase II&Normal&EWMA&Bodnar \& Schmid (2017)&Phase II& Normal(time series)&\centering\arraybackslash CUSUM\\
 Reynolds \& Stoumbos (2008)&Phase II&Normal& EWMA/Shewhart&Costa \& Neto (2017)& Phase II &Normal& \centering\arraybackslash Shewhart\\ 
 Riaz \& Does (2008)&Phase II&Normal&Shewhart&Fan et al. (2017)&Phase II&Normal&\centering\arraybackslash EWMA\\
 Zhang \& Chang (2008)&Phase II&Normal&EWMA&Gunaratne et al. (2017)& Phase II&Normal&\centering\arraybackslash EWMA\\
 Bodnar et al. (2009) &Phase II&Normal&EWMA/CUSUM&Huwang et al. (2017)&Phase II&Normal&\centering\arraybackslash EWMA\\
 Costa \& Machado (2009)&Phase II&Normal&Shewhart&Hwang (2017)&Phase II&Normal&\centering\arraybackslash -\\ 
 Machado et al. (2009a)&Phase II&Normal&Shewhart&Lee \& Khoo (2017a)&Phase II&Normal&\centering\arraybackslash Shewhart (synthetic)\\
 Machado et al. (2009b)&Phase II& Normal&Shewhart (syn)&Lee \& Khoo (2017b)&Phase II&Normal&\centering\arraybackslash Shewhart (synthetic)\\
 Mason et al. (2009)&Phase II&Normal&Shewhart& Mostajeran et al. (2017)&Phase II&(Non) Normal&\centering\arraybackslash Non-parametric\\
Memar \& Niaki (2009)&Phase II&Normal&EWMA& Osei-Aning et al. (2017)&Phase II&(Non) Normal&\centering\arraybackslash Shewhart\\
Niaki \& Memar (2009)&Phase II&Normal&EWMA& Saghir et al. (2017)&Phase I&(Non) Normal&\centering\arraybackslash Shewhart\\
 Zamba \& Hawkins (2009)&Phase II/change point&Normal&-& Gadre \& Kakade (2018)&Phase II&Normal&\centering\arraybackslash Shewhart (group runs)\\
 Mason et al. (2010)&Diagnosis&Normal&-&Haq \& Khoo (2018)&Phase II&(Non) Normal&\centering\arraybackslash Non-parametric\\ 
 Reynolds and Stoumbos (2010)&Phase II&Normal&EWMA/Shewhart& Lee \& Khoo (2018)&Phase II&Normal&\centering\arraybackslash Shewhart (double sampling)\\
Yen \& Shiau (2010)&Phase II&Normal&Shewhart& Machado et al. (2018)&Phase II&Normal&\centering\arraybackslash Shewhart\\
 Zhang et al. (2010)&Phase II&Normal&EWMA& Mingoti \& Pinto (2018)&Phase II&Normal&\centering\arraybackslash Shewhart\\
 Cheng \& Cheng (2011)&Phase II&Normal&-&Abdella et al. (2019)&Phase II&Normal&\centering\arraybackslash Shewhart (adaptive)\\
 Costa \& Machado (2011a)&Phase II&Normal&Shewhart&Alfaro \& Ortega (2019)&Phase II&Normal&\centering\arraybackslash EWMA \\
 Costa \& Machado (2011b)&Phase II&Normal&Shewhart&Haq \& Khoo (2019)&Phase II&Normal&\centering\arraybackslash EWMA/CUSUM (adaptive) \\
 Dogu \& Deveci-Kocakoc (2011)&Change point&Normal&Shewhart&Kim et al. (2019)&Phase II&Normal&\centering\arraybackslash Shewhart\\
 Maboudou-Tchao \&Hawkins (2011)&Phase II&Normal&EWMA&Liang et al. (2019)&Phase II&(Non) Normal&\centering\arraybackslash Non-parametric\\
 Memar \& Niaki (2011)&Phase II&Normal&EWMA&Li \& Tsung (2019)&Phase II&Normal&\centering\arraybackslash EWMA\\
 Reynolds \& Cho (2011)&Phase II&Normal&EWMA/Shewhart&Riaz et al. (2019)&Phase II&Normal&\centering\arraybackslash EWMA/CUSUM\\ 
 Zou et al. (2011)&Diagnosis&Normal&-&Tasias \& Nenes (2019)&Phase II&Normal&\centering\arraybackslash Shewhart (adaptive)\\
 Hung \& Chen (2012)&Diagnosis&Normal&-&Abdella et al. (2020)& Phase I&Normal&\centering\arraybackslash Shewhart (adaptive)\\
 Quinino et al. (2012)&Phase II&Normal&Shewhart&Ajadi  \& Zwetsloot (2020)&Phase II &Normal&\centering\arraybackslash Shewhart/EWMA\\
 Salehi et al. (2012)&Phase II&Normal&-&Bartzis \& Bersimis  (2020)&Phase II&Normal&\centering\arraybackslash Shewhart/EWMA\\
 Yeh et al. (2012)&Phase II&Normal&EWMA&Cabral Morais et al. (2020)&Misleading signals&Normal&\centering\arraybackslash Shewhart\\
 Yen et al. (2012)&Phase II&Normal&Shewhart&Fan et al. (2020)&Phase I&(Non) Normal&\centering\arraybackslash Non-parametric\\
 Costa \& Machado (2013)&Phase II&Normal&Shewhart&Koutras \& Soﬁkitou (2020)&Phase II &(Non) Normal&\centering\arraybackslash Non-parametric\\ 
 Dogu \& Deveci-Kocakoc (2013)&Change point&Normal&Shewhart&Ning \& Li (2020)&Phase II&Normal&\centering\arraybackslash EWMA\\ 
 Li et al. (2013)&Phase II&Normal&Shewhart&Osei-Aning \& Abbasi (2020)&Phase II&(Non) Normal&\centering\arraybackslash EWMA\\
 Maboudou-Tchao \& Agboto (2013)&Phase II&Normal&Shewhart&Sabahno et al. (2020a)&Phase II&Normal&\centering\arraybackslash Shewhart (adaptive)\\ 
 Maboudou-Tchao \& Diawara (2013)&Phase II&Normal&EWMA&Sabahno et al. (2020b)&Phase II&Normal&\centering\arraybackslash Shewhart (adaptive)\\
 Ramos et al. (2013)&Misleading signals&Normal&EWMA/Shewhart&Wang et al. (2020)&Diagnosis&Normal&\centering\arraybackslash -\\
& & &&&&&\\ 
 \hline
\end{tabular}
\end{sidewaystable}

\section{Conclusions and future research directions}
 In this paper, we review the existing literature on monitoring a covariance matrix of a multivariate process. The existing control charts are classified into four major groups where, for each group, the main ideas and results along with their benefits or drawbacks are briefly discussed. Most authors have focused on monitoring covariance matrices in situations where the number of rational subgroups is larger than the number of variables and a full rank estimate of the covariance matrix is available. However, there are many other situations where the literature on monitoring covariance matrices is limited. In what follows, we list the interesting future research problems relevant to the aforementioned work discussed in this paper. 
\begin{itemize}
\item[$\bullet$] As the number of control charts for monitoring covariance matrices is rapidly growing, there is a need to comparing methods to provide guidance for practitioners on how to choose a method among the existing control charts. As mentioned earlier in this paper, Vargas and Lagos (2007), Bartzis and Bertsimis (2020), and Ning and Li (2020) compared methods for monitoring covariance matrices. Recently, Ajadi and Zwetsloot (2020) performed a comparison in terms of Average Time to Signal (ATS) between some Phase II control charts  based on individual observations, (nonoverlapping) subgroups, and overlapping subgroups. In comparison to the usual (nonoverlapping) subgroups with fixed window, the subgroups are formed in a moving window in overlapping subgroups. Based on a simulation study, their recommendation is to use a multivariate chart with individual observations and if subgroups are used, overlapping subgroups are preferable. However, further comparisons among existing methods in both Phases I and II are still desirable. 
\item[$\bullet$ ] Misleading signals are common in simultaneous monitoring schemes for the process mean vector and the covariance matrix. For example when the process mean increases, a signal might show up in the chart for the covariance matrix.  Ramos et al. (2013) and Ramos et al. (2016) investigated the probabilities of misleading signals (PMS) for a few control chart methods such as Hotelling-$|\,\mathbf{S}\,|$  simultaneous scheme or simultaneous EWMA scheme chart. Recently, Cabral Morais et al. (2020) considered  three joint schemes for monitoring the mean vector and covariance matrix which all are multivariate extensions of univariate charts based on $\overline{X}$ and $S^2$ and calculated the PMS for them. However, this has not been yet studied for other types of control charts and more research is needed to better understand the PMS in other existing simultaneous control charts. 
\item[$\bullet$ ] A common problem in control charts is the low power for detecting small changes. The problem is often addressed by using CUSUM-type or EWMA-type control charts (see Woodall 1986 and Lucas and Saccucci 1990) rather than Shewhart-type charts, but a more recent approach is adding an adaptive feature, such as variable sample sizes or variable sampling intervals, to the control charts. Although adaptive charts are common in SPM, designing such control charts is almost new in the context of covariance monitoring. Haq and Khoo (2019), Tasias and Nenes (2019), and Sabahno et al. (2020a , b) are a few examples that recently considered applying adaptive features on some dispersion charts to achieve better performance over a range of shifts. More research is still needed, especially for those control charts whose adaptive versions have not been yet considered. See Perdikis and Psarakis (2019) as a recent review on multivariate adaptive control charts. 
\item[$\bullet$ ] The sparsity  feature in monitoring covariance matrices has been considered by some authors by employing  a penalized likelihood function.  The performance of these methods is highly dependent on the tuning parameter, but there is not much research on how to choose this parameter effectively.  In addition, as discussed in Shen et al. (2014), the joint monitoring of mean vector and covariance matrix under sparsity has not received much attention. 
\item[$\bullet$ ] Generally, monitoring procedures that are capable of distinguishing between mean shifts and covariance shifts are preferable.  Obviously, using separate charts is a solution, but there still remain problems to be solved when constructing these separate charts. Interpreting an out-of-control signal which includes the identification of the variables responsible for the shift and also diagnosing whether mean or covariance matrix has shifted should receive greater attention.
\item[$\bullet$ ] Effects of parameter estimation, as the main task of Phase I, on control chart properties is an important research subject in SPM area (Woodall and Montgomery 1999). As stated by Jones-Farmer et al. (2014), only a few studies have investigated the effect of estimation on the performance of control charts for covariance matrix. Addressing questions such as ``how well a chart performs if designed with estimates instead of exact parameters?" or ``what sample size is required in Phase I to ensure proper performance in Phase II?" is more challenging for covariance matrix monitoring as compared to monitoring a mean vector due to the large number of unknown parameters.
\item[$\bullet$ ] Using robust estimators instead of classical estimators can increase the performance of Phase I control charts under contaminated data. The literature is lacking on robust and easy-to-use multivariate control charts for Phase I monitoring, especially for the case of individual observations.
\item[$\bullet$ ] The ``curse of dimensionality" is increasingly encountered in industries such as biology, stock market analysis in finance, and wireless communication networks. Within the past 20 years, there has been a surge in statistical methods for high-dimensional data. See for example Jiang et al. (2012) and Srivastava et al. (2014). Using such techniques seems very promising for SPM. The work by Fan et al. (2020) is an example. Furthermore, in high-dimensional applications, the out-of-control conditions typically involve a small number of variables and consequently, combining a multivariate monitoring scheme with effective variable selection based methods would be yet another alternative for high-dimensional process monitoring. Interested readers are referred to Peres and Fogliatto (2018). 
\item[$\bullet$ ] In many of the methodologies presented in this paper, it is assumed that the underlying process follows a multivariate normal distribution. Little work has been done on dispersion charts with multivariate non-normal observations. To alleviate the effect of departure from multivariate normality, some distribution-free control schemes have been introduced for monitoring parameters of both univariate and multivariate processes. See Koutras and Triantafyllou (2020) for a recent overview. Nonparametric control charts enjoy greater robustness as compared with parametric control schemes. However, the existing work for monitoring a covariance matrix, to the best of the authors’ knowledge, is mainly concentrated on bivariate processes. There is a great need for nonparametric control charts for monitoring covariance matrices when two or more quality characteristics are present in the process. A similar gap exists for Phase I analysis, since developing nonparametric methods for Phase I is much more appropriate than for Phase II monitoring (Woodall 2017).
\item[$\bullet$ ] Monitoring covariance matrices of multivariate processes when the data are auto-correlated is another topic which needs more attention.  Chang and Zhang (2007) proposed a novel multivariate dynamic linear model (DLM) instead of the classical time series such as autoregressive integrated moving average (ARIMA) models to filter the autocorrelation in monitoring the covariance matrix of multivariate autocorrelated observations. In a recent attempt, Bodnar and Schmid (2017) considered modified MCUSUM control charts for detecting changes in the covariance structure of multivariate time series. More work is still needed for the cases where the assumption of independent observations in a multivariate process is violated.

\item[$\bullet$ ] The methods presented in this paper are data driven in a sense that data are collected and used for process monitoring; therefore the methods are ``noncausal" in nature. However, adding more causal-oriented structure or a priori knowledge to the process monitoring approaches could increase their effectiveness. Recent attempts try to include process-specific structure into the monitoring procedures which may bring benefits not only for detection but also for diagnosis and troubleshooting activities. This is not in the scope of this paper, but a helpful review on this topic is given recently by Reis et al. (2019). 
\end{itemize}

\section*{References}
\begin{enumerate}
\item[] Abdella, G. M., Kim, J., Kim, S., Al-Khalifa, K. N., Jeong, M. K., Hamouda, A. M., \& Elsayed, E. A. (2019). An adaptive thresholding-based process variability monitoring. {\it Journal of Quality Technology, 51}, 1-14.
\item[] Abdella, G. M., Maleki, M. R., Kim, S., Al-Khalifa, K. N., \& Hamouda, A. M. S. (2020). Phase-I monitoring of high-dimensional covariance matrix using an adaptive thresholding LASSO rule. {\it  Computers \& Industrial Engineering}, 106465.
\item[] Ajadi, J. O., \& Zwetsloot, I. M. (2020). Should observations be grouped for effective monitoring of multivariate process variability? {\it Quality and Reliability Engineering International, 36}, 1005-1027.
\item[] Alfaro, J. L., \& Ortega, J. F. (2019). A new multivariate variability control chart based on a covariance matrix combination. {\it Applied Stochastic Models in Business and Industry, 35}, 823-836.
\item[] Alt FB. (1985). Multivariate quality control.{\it The Encyclopedia of Statistical Sciences}, Kotz S, Johnson NL, Read CR (eds.).Wiley: New York, 110-122.
\item[] Alt, F. B., \& Smith, N. D. (1988). Multivariate process control. In: Krishnaiah, P. R., Rao, C. R., eds. {\it Handbook of Statistics}, vol. 7, New York: Elsevier, 331–351.
\item[] Bartzis, G., \& Bersimis, S. (2020). Performance comparisons of bivariate dispersion control charts. {\it Communications in Statistics: Case Studies, Data Analysis and Applications, 6}, 69-85.
\item[] Bersimis, S., Psarakis, S., \& Panaretos, J. (2007). Multivariate statistical process control charts: an overview. {\it Quality and Reliability engineering international, 23}, 517-543.
\item[] Bodnar, O., Bodnar, T., \& Okhrin, Y. (2009). Surveillance of the covariance matrix based on the properties of the singular Wishart distribution. {\it Computational statistics \& data analysis}, 53, 3372-3385.
\item[] Bodnar, O., \& Schmid, W. (2017). CUSUM control schemes for monitoring the covariance matrix of multivariate time series. {\it Statistics, 51}, 722-744.
\item[] Cabral Morais, M., Schmid, W., Ferreira Ramos, P., Lazariv, T., \& Pacheco, A. (2020). Misleading signals in joint schemes for the mean vector and covariance matrix.{\it Quality and Reliability Engineering International, 36}, 642-651.
\item[] Cai, T., \& Liu, W. (2011). Adaptive thresholding for sparse covariance matrix estimation. {\it Journal of the American Statistical Association, 106}, 672-684.
\item[] Chakraborti, S., Van der Laan, P., \& Bakir, S. T. (2001). Nonparametric control charts: an overview and some results. {\it Journal of Quality Technology, 33}, 304-315.
\item[] Chang, S. I., \& Zhang, K. (2007). Statistical process control for variance shift detections of multivariate autocorrelated processes. {\it Quality Technology \& Quantitative Management, 4}, 413-435.
\item[] Cheng, C. S., \& Cheng, H. P. (2008). Identifying the source of variance shifts in the multivariate process using neural networks and support vector machines. {\it Expert Systems with Applications, 35}, 198-206.
\item[] Cheng, C. S., \& Cheng, H. P. (2011). Using neural networks to detect the bivariate process variance shifts pattern. {\it Computers \& Industrial Engineering, 60}, 269-278.
\item[] Cheng, C. S., \& Lee, H. T. (2016). Diagnosing the variance shifts signal in multivariate process control using ensemble classifiers. {\it Journal of the Chinese Institute of Engineers, 39}, 64-73.
\item[] Costa, A. F. B., Machado, M. A. G. (2008). A new chart for monitoring the covariance matrix of bivariate processes. {\it Communications in Statistics - Simulation and Computation, 37}, 1453-1465.
\item[] Costa, A. F. B., Machado, M. A. G. (2009). A new chart based on sample variances for monitoring the covariance matrix of multivariate processes. {\it The International Journal of Advanced Manufacturing Technology, 41}, 770-779.
\item[] Costa, A. F. B., \& Machado, M. A. G. (2011a). A control chart based on sample ranges for monitoring the covariance matrix of the multivariate processes. {\it Journal of Applied Statistics, 38}, 233-245.
\item[] Costa, A. F. B., \& Machado, M. A. G. (2011b). Monitoring the mean vector and the covariance matrix of multivariate processes with sample means and sample ranges. {\it Producao, 21}, 197-208.
\item[] Costa, A. F., \& Machado, M. A. (2013). A single chart with supplementary runs rules for monitoring the mean vector and the covariance matrix of multivariate processes.{\it Computers \& Industrial Engineering, 66}, 431-437.
\item[] Costa, A. F. B., \& Neto, A. F. (2017). The S chart with variable charting statistic to control bi and trivariate processes. {\it Computers \& Industrial Engineering, 113}, 27-34.
\item[] Djauhari, M. A., Mashuri, M., \& Herwindiati, D. E. (2008). Multivariate process variability monitoring. {\it Communications in Statistics-Theory and Methods, 37}, 1742-1754.
\item[] Djauhari, M. A., Sagadavan, R., \& Li, L. S. (2016). Monitoring multivariate process variability when sub-group size is small. {\it Quality Engineering, 28}, 429-440.
\item[] Dogu, E. (2015). Identifying the time of a step change with multivariate single control charts. {\it Journal of Statistical Computation and Simulation, 85}, 1529-1543.
\item[] Dogu, E., \&  Deveci-Kocakoc, I. (2011). Estimation of change point in generalized variance control chart. {\it Communications in Statistics-Simulation and Computation, 40}, 345-363.
\item[] Dogu, E., \&  Deveci-Kocakoc, I. D. (2013). A multivariate change point detection procedure for monitoring mean and covariance simultaneously. {\it Communications in Statistics-Simulation and Computation, 42}, 1235-1255.
\item[] Fan, J., Shu, L., Yang, A., \& Li, Y. (2020). Phase I analysis of high-dimensional covariance matrices based on sparse leading eigenvalues. {\it Journal of Quality Technology}, 1-14.
\item[] Fan, J., Shu, L., Zhao, H., \& Yeung, H. (2017). Monitoring multivariate process variability via eigenvalues. {\it Computers \& Industrial Engineering, 113}, 269-281.
\item[] Gadre, M. P. (2014). A multivariate group runs control chart for process dispersion. {\it Communications in Statistics-Simulation and Computation, 43}, 813-837.
\item[] Gadre, M. P., \& Kakade, V. C. (2018). Two group inspection-based control charts for dispersion matrix. {\it Communications in Statistics-Simulation and Computation, 47}, 1652-1669.
\item[] Ghute, V. B., Shirke, D. T. (2008). A multivariate synthetic control chart for process dispersion. {\it Quality Technology and Quantitative Management, 5}, 271-288.
\item[] Gunaratne, N. G. T., Abdollahian, M. A., Huda, S., \& Yearwood, J. (2017). Exponentially weighted control charts to monitor multivariate process variability for high dimensions. {\it International Journal of Production Research, 55}, 4948-4962.
\item[] Haq, A., \& Khoo, M. B. (2018). A new non-parametric multivariate EWMA sign control chart for monitoring process dispersion. {\it Communications in Statistics-Theory and Methods, 48}, 3703-3716.
\item[] Haq, A., \& Khoo, M. B. (2019). Multivariate process dispersion monitoring without subgrouping. {\it Journal of Applied Statistics}, 1-24.
\item[] Hawkins, D. M., \& Maboudou-Tchao, E. M. (2008). Multivariate exponentially weighted moving covariance matrix. {\it Technometrics, 50}, 155-166.
\item[] Healy, J. D. (1987). A note on multivariate CUSUM procedures. {\it Technometrics, 29}, 409-412.
\item[] Hotelling, H. (1947). Multivariate Quality Control, illustrated by the air testing of sample bombsights. {\it Techniques of Statistical Analysis}, Eisenhart C, Hastay MW, Wallis WA (eds.). McGraw-Hill: New York, 111-184.
\item[] Hung, H., \& Chen, A. (2012). Test of covariance changes without a large sample and its application to fault detection and classification. {\it Journal of Process Control, 22}, 1113-1121.
\item[] Huwang, L., Lin, P. C., Chang, C. H., Lin, L. W., \& Tee, Y. S. (2017). An EWMA chart for monitoring the covariance matrix of a multivariate process based on dissimilarity index. {\it Quality and Reliability Engineering International, 33}, 2089-2104.
\item[] Huwang L, Yeh AB, Wu CV. (2007). Monitoring multivariate process variability for individual observations. {\it Journal of Quality Technology, 39}, 258-278.
\item[] Hwang, W. Y. (2017). Chi-Square quantile-based multivariate variance monitoring for individual observations. {\it Communications in Statistics-Simulation and Computation, 46}, 1-18.
\item[] Jackson, J. E. (1985) Multivariate quality control. {\it Communications in Statistics - Theory and Methods, 14}, 2657-2688.
\item[] Jiang, D., Jiang, T., \& Yang, F. (2012). Likelihood ratio tests for covariance matrices of high-dimensional normal distributions. {\it Journal of Statistical Planning and Inference, 142}, 2241-2256.
\item[] Jones-Farmer, L. A., Woodall, W. H., Steiner, S. H., Champ, C. W. (2014). An overview of phase I analysis for process improvement and monitoring. {\it Journal of Quality Technology, 46}, 265-280.
\item[] Kim, J., Abdella, G. M., Kim, S., Al-Khalifa, K. N., \& Hamouda, A. M. (2019). Control charts for variability monitoring in high-dimensional processes. {\it Computers \& Industrial Engineering, 130}, 309-316.
\item[] Koutras, M. V., \& Triantafyllou, I. S. (Eds.). (2020). Distribution-free methods for statistical process monitoring and control. {\it Springer Nature.}
\item[] Koutras, M. V., \& Sofikitou, E. M. (2020). Bivariate semiparametric control charts for simultaneous monitoring of process mean and variance. {\it Quality and Reliability Engineering International, 36}, 447-473.
\item[] Ledoit, O., \& Wolf, M. (2002). Some hypothesis tests for the covariance matrix when the dimension is large compared to the sample size. {\it The Annals of Statistics, 30}, 1081-1102.
\item[] Lee, M. H., \& Khoo, M. B. (2015). Multivariate synthetic $|S|$ control chart with variable sampling interval. {\it Communications in Statistics-Simulation and Computation, 44}, 924-942.
\item[] Lee, M. H., \& Khoo, M. B. (2017a). Combined synthetic and $|S|$ chart for monitoring process dispersion. {\it Communications in Statistics-Simulation and Computation, 46}, 5698-5711.
\item[] Lee, M. H., \& Khoo, M. B. (2017b). Optimal designs of multivariate synthetic $|S|$ control chart based on median run length. {\it Communications in Statistics-Theory and Methods, 46}, 3034-3053.
\item[] Lee, M. H., \& Khoo, M. B. (2018). Double sampling $|S|$ control chart with variable sample size and variable sampling interval. {\it Communications in Statistics-Simulation and Computation, 47}, 615-628.
\item[] Li, Z., Dai, Y., \& Wang, Z. (2014). Multivariate change point control chart based on data depth for phase I analysis. {\it Communications in Statistics-Simulation and Computation, 43}, 1490-1507.
\item[] Li, Z., \& Tsung, F. (2019). A control scheme for monitoring process covariance matrices with more variables than observations. {\it Quality and Reliability Engineering International, 35}, 351-367.
\item[] Li, B., Wang, K., \& Yeh, A. B. (2013). Monitoring the covariance matrix via penalized likelihood estimation. {\it IIE Transactions, 45}, 132-146.
\item[] Liang, W., Xiang, D., Pu, X., Li, Y., \& Jin, L. (2019). A robust multivariate sign control chart for detecting shifts in covariance matrix under the elliptical directions distributions. {\it Quality Technology \& Quantitative Management, 16}, 113-127.
\item[] Lowry, C. A., \& Montgomery, D. C. (1995). A review of multivariate control charts. {\it IIE transactions, 27}, 800-810.
\item[] Lucas, J. M., \& Saccucci, M. S. (1990). Exponentially weighted moving average control schemes: Properties and enhancements. {\it Technometrics, 32}, 1-12.
\item[] Maboudou-Tchao, E. M., \& Agboto, V. (2013). Monitoring the covariance matrix with fewer observations than variables. {\it Computational Statistics \& Data Analysis, 64}, 99-112.
\item[] Maboudou-Tchao, E. M., \& Diawara, N. (2013). A lasso chart for monitoring the covariance matrix. {\it Quality technology \& quantitative management, 10}, 95-114.
\item[] Maboudou-Tchao, E.M., \& Hawkins, D.M., (2011) . Self-starting multivariate control charts for location and scale. {\it Journal of Quality Technology, 43}, 113-126.
\item[] Machado, M. A., \& Costa, A. F. (2008). The double sampling and the EWMA charts based on the sample variances. {\it International Journal of Production Economics, 114}, 134-148.
\item[] Machado, M. A., Costa, A. F., \& Marins, F. A. (2009a). Control charts for monitoring the mean vector and the covariance matrix of bivariate processes. {\it The International Journal of Advanced Manufacturing Technology, 45}, 772-785.
\item[] Machado, M. A., Costa, A. F., \& Rahim, M. A. (2009b). The synthetic control chart based on two sample variances for monitoring the covariance matrix. {\it Quality and Reliability Engineering International, 25}, 595-606.
\item[] Machado, M. A. G., Ho, L. L., \& Costa, A. F. B. (2018). Attribute control charts for monitoring the covariance matrix of bivariate processes. {\it Quality and Reliability Engineering International, 34}, 257-264.
\item[] Mason, R. L., Chou, Y. M., \& Young, J. C. (2009). Monitoring variation in a multivariate process when the dimension is large relative to the sample size. {\it Communications in Statistics-Theory and Methods, 38}, 939-951.
\item[] Mason, R. L., Chou, Y. M., \& Young, J. C. (2010). Decomposition of scatter ratios used in monitoring multivariate process variability. {\it Communications in Statistics-Theory and Methods, 39}, 2128-2145.
\item[] Memar, A. O., \& Niaki, S. T. A. (2009). New control charts for monitoring covariance matrix with individual observations. {\it Quality and Reliability Engineering International, 25}, 821-838.
\item[] Memar, A. O., \& Niaki, S. T. A. (2011). Multivariate variability monitoring using EWMA control charts based on squared deviation of observations from target. {\it Quality and Reliability Engineering International, 27}, 1069-1086.
\item[] Mingoti, S. A., \& Pinto, L. P. (2018). Step-down diagnostic analysis for monitoring the covariance matrix of bivariate normal processes. {\it Communications in Statistics-Simulation and Computation, 48}, 2615-2624.
\item[] Mostajeran, A., Iranpanah, N., \& Noorossana, R. (2017). An evaluation of the multivariate dispersion charts with estimated parameters under non-normality. {\it Applied Stochastic Models in Business and Industry, 33}, 694-716.
\item[] Niaki, S. T. A., \& Memar, A. O. (2009). A new statistical process control method to monitor and diagnose bivariate normal mean vectors and covariance matrices simultaneously. {\it The International Journal of Advanced Manufacturing Technology, 43}, 964-981.
\item[] Ning, X., \& Li, P. (2020). A simulation comparison of some distance-based EWMA control charts for monitoring the covariance matrix with individual observations. {\it Quality and Reliability Engineering International, 36}, 50-67.
\item[] Osei-Aning, R., \& Abbasi, S. A. (2020). Efficient bivariate EWMA charts for monitoring process dispersion. {\it Quality and Reliability Engineering International, 36}, 247-267.
\item[] Osei-Aning, R., Abbasi, S. A., \& Riaz, M. (2017). Bivariate Dispersion Control Charts for Monitoring Non-Normal Processes. {\it Quality and Reliability Engineering International, 33}, 515-529.
\item[] Perdikis, T., \& Psarakis, S. (2019). A survey on multivariate adaptive control charts: Recent developments and extensions. {\it Quality and Reliability Engineering International, 35}, 1342-1362.
\item[] Peres, F. A. P., \& Fogliatto, F. S. (2018). Variable selection methods in multivariate statistical process control: A systematic literature review. {\it Computers \& Industrial Engineering, 115}, 603-619.
\item[] Quinino, R., Costa, A., \& Lee Ho, L. (2012). A single statistic for monitoring the covariance matrix of bivariate processes. {\it Quality Engineering, 24}, 423-430.
\item[] Ramos, P. F., Morais, M. C., Pacheco, A., \& Schmid, W. (2013). Misleading signals in simultaneous schemes for the mean vector and covariance matrix of a bivariate process. In {\it Recent developments in modeling and applications in statistics}. Springer Berlin Heidelberg, pp 225-235.
\item[] Ramos, P. F., Morais, M. C., Pacheco, A., \& Schmid, W. (2016). On the misleading signals in simultaneous schemes for the mean vector and covariance matrix of multivariate iid output. {\it Statistical Papers, 57}, 471-498.
\item[] Reis, M. S., Gins, G., \& Rato, T. J. (2019). Incorporation of process-specific structure in statistical process monitoring: A review. {\it Journal of Quality Technology, 51}, 407-421.
\item[] Reynolds, MR Jr, \& Cho, GY (2006) Multivariate Control Charts for Monitoring the Mean Vector and Covariance Matrix. {\it Journal of Quality Technology, 38}, 230-253.
\item[] Reynolds Jr, M. R., \& Cho, G. Y. (2011). Multivariate control charts for monitoring the mean vector and covariance matrix with variable sampling intervals. {\it Sequential Analysis, 30}, 1-40.
\item[] Reynolds Jr, M. R., \& Kim, K. (2007). Multivariate control charts for monitoring the process mean and variability using sequential sampling. {\it Sequential Analysis, 26}, 283-315.
\item[] Reynolds MR, \& Stoumbos ZG. (2008). Combinations of multivariate Shewhart and MEWMA control charts for monitoring the mean vector and covariance matrix. {\it Journal of Quality Technology, 40}, 381-393.
\item[] Reynolds Jr, M. R., \& Stoumbos, Z. G. (2010). Multivariate monitoring of the process mean and variability using combinations of Shewhart and MEWMA control charts. {\it Frontiers in Statistical Quality Control, 9}, 37-54.
\item[] Riaz, M., Ajadi, J. O., Mahmood, T., \& Abbasi, S. A. (2019). Multivariate mixed EWMA-CUSUM control chart for monitoring the process variance-covariance matrix. {\it IEEE Access, 7}, 100174-100186.
\item[] Riaz, M., \& Does, R. J. (2008). An alternative to the bivariate control chart for process dispersion. {\it Quality Engineering, 21}, 63-71.
\item[] Sabahno, H., Castagliola, P., \& Amiri, A. (2020a). A variable parameters multivariate control chart for simultaneous monitoring of the process mean and variability with measurement errors. {\it Quality and Reliability Engineering International, 36}, 1161-1196.
\item[] Sabahno, H., Amiri, A., \& Castagliola, P. (2020b). A new adaptive control chart for the simultaneous monitoring of the mean and variability of multivariate normal processes. {\it Computers \& Industrial Engineering}, 106524.
\item[] Saghir, A. (2015). The bivariate dispersion control charts for non-normal processes. {\it International Journal of Production Research, 53}, 1964-1979.
\item[] Saghir, A., Chakraborti, S., \& Ahmad, I. (2017). On the performance of phase I bivariate dispersion charts to non-Normality. {\it Quality and Reliability Engineering International, 33}, 637-656.
\item[] Saghir, A., Khan, Y. A., \& Chakraborti, S. (2016). The phase I dispersion charts for bivariate process monitoring. {\it Quality and Reliability Engineering International, 32}, 1807-1823.
\item[] Salehi, M., Kazemzadeh, R. B.,\& Salmasnia, A. (2012). On line detection of mean and variance shift using neural networks and support vector machine in multivariate processes. {\it Applied Soft Computing, 12}, 2973-2984.
\item[] Shen, X., Tsung, F., \& Zou, C. (2014). A new multivariate EWMA scheme for monitoring covariance matrices. {\it International Journal of Production Research, 52}, 2834-2850.
\item[] Srivastava, M. S., Yanagihara, H., \& Kubokawa, T. (2014). Tests for covariance matrices in high dimension with less sample size. {\it Journal of Multivariate Analysis, 130}, 289-309.
\item[] Sullivan, J. H., Stoumbos, Z. G., Mason, R. L., \& Young, J. C. (2007). Step-Down Analysis for Changes in the Covariance Matrix and Other Parameters. {\it Journal of Quality Technology, 39}, 66-84.
\item[] Surtihadi, J., Raghavachari, M., \& Runger, G. (2004). Multivariate control charts for process dispersion. {\it International Journal of Production Research, 42}, 2993-3009.
\item[] Tasias, K. A., \& Nenes, G. (2019). Monitoring location and scale of multivariate processes subject to a multiplicity of assignable causes. {\it Quality Technology \& Quantitative Management,} 1-17.
\item[] Variyath, A. M., \& Vattathoor, J. (2014). Robust control charts for monitoring process variability in phase I multivariate individual observations. {\it Quality and Reliability Engineering International, 30}, 795-812.
\item[] Vargas, N. J. A., \& Lagos, C. J. (2007). Comparison of multivariate control charts for process dispersion. {\it Quality Engineering, 19}, 191-196.
\item[] Wang, B., Xu, F., \& Shu, L. (2020). A Bayesian approach to diagnosing covariance matrix shifts. {\it Quality and Reliability Engineering International, 36}, 736-752.
\item[] Wang, K., Yeh, A. B., \& Li, B. (2014). Simultaneous monitoring of process mean vector and covariance matrix via penalized likelihood estimation. {\it Computational Statistics \& Data Analysis, 78}, 206-217.
\item[] Woodall, W. H. (1986). The design of CUSUM quality control charts. {\it Journal of Quality Technology, 18}, 99-102.
\item[] Woodall, W. H. (2000). Controversies and contradictions in statistical process control. {\it Journal of Quality Technology, 32}, 341-350.
\item[] Woodall, W. H. (2017). Bridging the gap between theory and practice in basic statistical process monitoring. {\it Quality Engineering, 29}, 2-15.
\item[] Woodall, W. H., \& Montgomery, D. C. (1999). Research issues and ideas in statistical process control. {\it Journal of Quality Technology, 31}, 376-386.
\item[] Yeh, A. B., Lin, D. K., \& McGrath, R. N. (2006). Multivariate control charts for monitoring covariance matrix: a review. {\it Quality Technology \& Quantitative Management, 3}, 415-436.
\item[] Yeh, A. B., Li, B., \& Wang, K. (2012). Monitoring multivariate process variability with individual observations via penalised likelihood estimation. {\it International Journal of Production Research, 50}, 6624-6638.
\item[] Yen, C. L., Shiau, J. J. H., \& Yeh, A. B. (2012). Effective control charts for monitoring multivariate process dispersion. {\it Quality and Reliability Engineering International, 28}, 409-426.
\item[] Yen, C. L., \& Shiau, J. J. H. (2010). A multivariate control chart for detecting increases in process dispersion. {\it Statistica Sinica, 20}, 1683-1707.
\item[] Zamba K.D., \& Hawkins, D. M. (2009) A Multivariate Change-Point Model for Change in Mean Vector and/or Covariance Structure. {\it Journal of Quality Technology, 41}, 285-303.
\item[] Zhang, G., \& Chang, S. I. (2008). Multivariate EWMA control charts using individual observations for process mean and variance monitoring and diagnosis. {\it International Journal of Production Research, 46}, 6855-6881.
\item[] Zhang, C., Chen, N., \& Zou, C. (2016). Robust multivariate control chart based on goodness-of-fit test. {\it Journal of Quality Technology, 48}, 139.
\item[] Zhang, J., Li, Z., \& Wang, Z. (2010). A multivariate control chart for simultaneously monitoring process mean and variability. {\it Computational Statistics \& Data Analysis, 54}, 2244-2252.
\item[] Zhou, M., Zi, X., Geng, W., \& Li, Z. (2015). A distribution-free multivariate change-point model for statistical process control. {\it Communications in Statistics-Simulation and Computation, 44}, 1975-1987.
\item[] Zou, C., Jiang, W., \& Tsung, F. (2011). A LASSO-based diagnostic framework for multivariate statistical process control. {\it Technometrics, 53}, 297-309.

\end{enumerate}
\end{document}